{}
{}

\documentclass[12pt,a4paper]{article}

\newif\ifpublic\publictrue
\newif\ifniklas\niklasfalse

\usepackage{ifpdf}
\ifniklas\ifpdf\publictrue\else\publicfalse
\PassOptionsToPackage{hypertex}{hyperref}
\PassOptionsToPackage{draft}{graphicx}
\fi\fi

\PassOptionsToPackage{bookmarks=true,hyperfigures=true}{hyperref}
\usepackage{amsmath,amssymb}
\ifpublic\else\usepackage{showkeys}\fi
\usepackage{hyperref}
\usepackage{graphicx}
\usepackage[nosort]{cite}
\usepackage[bulletsep]{collref}

\ifniklas

\def\showkeysrefformat#1{{\normalfont\tiny\ttfamily#1}}
\makeatletter
\def\SK@@ref#1>#2\SK@{%
 {\@inlabelfalse\leavevmode\vbox to\z@{%
 \vss\SK@refcolor\rlap{\vrule\raise .75em%
  \hbox{\showkeysrefformat{#2}}}}}}
\makeatother
\fi

\usepackage[a4paper,text={450pt,650pt},centering]{geometry}

\expandafter\def\expandafter\bfseries\expandafter{\bfseries\ifmmode\else\boldmath\fi}
\expandafter\def\expandafter\mdseries\expandafter{\mdseries\ifmmode\else\unboldmath\fi}
\expandafter\def\expandafter\normalfont\expandafter{\normalfont\ifmmode\else\unboldmath\fi}

\allowdisplaybreaks[3]

\numberwithin{equation}{section}

\usepackage[font=small,labelfont=bf,width=0.85\textwidth]{caption}

\providecommand{\href}[2]{#2}
\providecommand{\hypersetup}[1]{}
\providecommand{\texorpdfstring}[2]{#1}

\hypersetup{plainpages=false}
\hypersetup{pdfpagemode=UseNone}
\hypersetup{bookmarksnumbered=true}
\hypersetup{pdfstartview=FitH}
\hypersetup{colorlinks=false}
\hypersetup{citebordercolor={.5 1 .5}}
\hypersetup{urlbordercolor={.5 1 1}}
\hypersetup{linkbordercolor={1 .7 .7}}

\makeatletter
\let\@keywords\@empty
\let\@subject\@empty
\providecommand{\keywords}[1]{\gdef\@keywords{#1}}
\providecommand{\subject}[1]{\gdef\@subject{#1}}
\def\thetitle{\@title}
\def\theauthor{\@author}
\def\thesubject{\@subject}
\def\thedate{\@date}
\def\thekeywords{\@keywords}
\AtBeginDocument{
\hypersetup{pdftitle={\thetitle}}%
\hypersetup{pdfauthor={\theauthor}}%
\hypersetup{pdfsubject={\thesubject}}%
\hypersetup{pdfkeywords={\thekeywords}}%
}
\makeatother

\usepackage{fixmath}

\newcommand{\gen}[1]{\mathrm{#1}}

\newcommand{\copro}{\mathrm{\Delta}}
\newcommand{\counit}{\varepsilon}
\newcommand{\coproop}{\widetilde{\copro}}
\newcommand{\antipode}{\mathrm{S}}
\newcommand{\rmat}{\mathcal{R}}

\newcommand{\order}[1]{\mathcal{O}(#1)}

\ifx\genfrac\sdflkaj\else\fi
\newcommand{\sfrac}[2]{{\textstyle\frac{#1}{#2}}}
\newcommand{\half}{\sfrac{1}{2}}

\newcommand{\quarter}{\sfrac{1}{4}}

\newcommand{\indup}[1]{_{\mathrm{#1}}}

\newcommand{\matr}[2]{\left(\begin{array}{#1}#2\end{array}\right)}
\newcommand{\tinymatr}[2]{\left(\text{\tiny\arraycolsep0.3em$\begin{array}{#1}#2\end{array}$}\right)}

\newcommand{\bigbrk}[1]{\bigl(#1\bigr)}

\newcommand{\bigcomm}[2]{\big[#1,#2\big]}
\newcommand{\comm}[2]{[#1,#2]}

\newcommand{\acomm}[2]{\{#1,#2\}}
\newcommand{\bigacomm}[2]{\big\{#1,#2\big\}}
\newcommand{\scomm}[2]{[#1,#2\}}

\newcommand{\set}[1]{\{#1\}}

\newcommand{\alg}[1]{\mathfrak{#1}}

\newcommand{\halg}[1]{\mathcal{#1}}

\newcommand{\nln}{\nonumber\\}

\def\[{\begin{equation}}
\def\]{\end{equation}}

\makeatletter
\def\mr@ignsp#1 {\ifx\:#1\@empty\else #1\expandafter\mr@ignsp\fi}%
\newcommand{\multiref}[1]{\begingroup
\xdef\mr@no@sparg{\expandafter\mr@ignsp#1 \: }%
\def\mr@comma{}%
\@for\mr@refs:=\mr@no@sparg\do{\mr@comma\def\mr@comma{,}\ref{\mr@refs}}%
\endgroup}
\renewcommand{\eqref}[1]{(\multiref{#1})}
\makeatother

\makeatletter
\newcommand{\namedref}[2]{\hyperref[#2]{#1~\ref*{#2}}}
\newcommand{\secref}{\@ifstar{\namedref{Section}}{\namedref{Sec.}}}
\newcommand{\appref}{\@ifstar{\namedref{Appendix}}{\namedref{App.}}}
\newcommand{\tabref}{\@ifstar{\namedref{Table}}{\namedref{Tab.}}}
\newcommand{\figref}{\@ifstar{\namedref{Figure}}{\namedref{Fig.}}}
\makeatother

\makeatletter
\newsavebox{\apb@box}\newlength{\apb@width}
\newcommand{\autoparbox}[2][c]{\sbox{\apb@box}{#2}%
 \settowidth{\apb@width}{\usebox{\apb@box}}%
 \parbox[#1]{\apb@width}{\usebox{\apb@box}}}
\newcommand{\includegraphicsbox}[2][]{\autoparbox{\includegraphicsex[#1]{#2}}}
\makeatother

\RequirePackage{verbatim}

\makeatletter
\newwrite\bibinl@out
\newenvironment{bibtex}[1][\jobname]{%
  \immediate\openout\bibinl@out #1.bib
  \immediate\write\bibinl@out{\@percentchar generated from `\jobname' starting line \the\inputlineno^^J}%
  \def\verbatim@processline{\immediate\write\bibinl@out{\the\verbatim@line}}%
  \@bsphack\let\do\@makeother\dospecials\catcode`\^^M\active\verbatim@start
}%
{\immediate\closeout\bibinl@out\@esphack}
\makeatother

\RequirePackage{verbatim}

\makeatletter
\newwrite\mpi@out
\def\mpi@write#1{\immediate\write\mpi@out{#1}}
\immediate\openout\mpi@out\jobname.mp
\mpi@write{\@percentchar generated from `\jobname'}%
\AtEndDocument{\immediate\closeout\mpi@out}

\newcommand{\mpi@putlineno}{%
  \mpi@write{\@percentchar---------------------------------------}%
  \mpi@write{\@percentchar l.\the\inputlineno}%
}

\newcommand{\mpi@verbatim}{
  \@bsphack
  \let\do\@makeother\dospecials
  \catcode`\^^M\active
  \def\verbatim@processline{\mpi@write{\the\verbatim@line}}%
  \verbatim@start
}

\newenvironment{mpostcmd}{%
  \mpi@putlineno%
  \mpi@verbatim%
}%
{\mpi@write{}\@esphack}

\newenvironment{mpostfile}[1]{%
  \mpi@putlineno%
  \mpi@write{filenametemplate "#1";}%
  \mpi@write{beginfig(0)}%
  \mpi@verbatim%
}%
{\mpi@write{endfig;}\@esphack}

\newcommand{\includegraphicsex}[2][]{%
  \xdef\mpi@tmp{#2}%
  \IfFileExists{\mpi@tmp}%
    {\includegraphics[#1]{\mpi@tmp}}%
    {\textbf{??}\typeout{file \mpi@tmp{} missing}}%
}

\makeatother

\newcommand{\arxivlink}[1]{\href{http://arxiv.org/abs/#1}{arxiv:#1}}

\begin{mpostcmd}
prologues:=2;

verbatimtex 
\documentclass{article}
\usepackage{amsfonts,amssymb}
\usepackage{latexsym}
\newcommand{\gen}[1]{\mathrm{#1}}
\begin{document}
etex

picture copyrightline,copyleftline;
copyrightline := btex \copyright\ \textsf{2011 Niklas Beisert} etex; copyleftline := btex $\circledast$ etex;

def putcopyspace = 
label.bot(btex \vphantom{gA} etex scaled 0.1, lrcorner(currentpicture));
enddef;
def putcopy = 
label.ulft(copyrightline scaled 0.1, lrcorner(currentpicture)) withcolor 0.9white;
label.urt(copyleftline scaled 0.1, llcorner(currentpicture)) withcolor 0.9white;
currentpicture:=currentpicture shifted (10.5cm,14cm);
enddef;

ahlength:=7pt;

dotsize:=7pt;
crosssize:=4pt;
dashdist:=6pt;
dotdist:=3pt;
xu:=1cm;
yu:=1cm;

wiggly_len := 3mm; 
wiggly_slope := 60;
curly_len := 3mm;
zigzag_len := 2mm;
zigzag_width := 1.5pt;
double_size := 2pt;
double_width := 1pt;

def colgrey=withcolor (0.5,0.5,0.5) enddef;

def pensize(expr s)=withpen pencircle scaled s enddef;
def thinpen=pensize(0.1pt) enddef;
def mediumpen=pensize(0.2pt) enddef;
def normalpen=pensize(0.4pt) enddef;
def thickpen=pensize(0.6pt) enddef;
def fatpen=pensize(0.8pt) enddef;

def dashdots=dashed withdots scaled (dotdist/5pt) enddef;
def dasheven=dashed evenly scaled (dashdist/6pt) enddef;
def dash(expr a,b)=dashed dashpattern(on (a*b) off ((1-a)*b)) enddef;

def onwhite(expr p)=unfill bbox p; draw p; enddef;

vardef pixlen (expr p, n) =
  for k=1 upto length(p): + segment_pixlen (subpath (k-1,k) of p, n) endfor
enddef;
vardef segment_pixlen (expr p, n) =
  for k=1 upto n: + abs (point k/n of p - point (k-1)/n of p) endfor
enddef;

vardef wiggly expr p_arg =
 save wpp;
 numeric wpp;
 wpp = ceiling (pixlen (p_arg, 10) / wiggly_len) / length p_arg;
 for k=0 upto wpp*length(p_arg) - 1:
  point k/wpp of p_arg
       {direction k/wpp of p_arg rotated wiggly_slope} ..
  point (k+.5)/wpp of p_arg
       {direction (k+.5)/wpp of p_arg rotated - wiggly_slope} ..
 endfor
 if cycle p_arg: cycle else: point infinity of p_arg fi
enddef;
vardef curly expr p =
 save cpp;
 numeric cpp;
 cpp := ceiling (pixlen (p, 10) / curly_len) / length p;
 if cycle p:
   for k=0 upto cpp*length(p) - 1:
     point (k+.33)/cpp of p
           {direction (k+.33)/cpp of p rotated 90} ..
     point (k-.33)/cpp of p
           {direction (k-.33)/cpp of p rotated -90} ..
   endfor
   cycle
 else:
   point 0 of p
         {direction 0 of p rotated -90} ..
   for k=1 upto cpp*length(p) - 1:
     point (k+.33)/cpp of p
           {direction (k+.33)/cpp of p rotated 90} ..
     point (k-.33)/cpp of p
           {direction (k-.33)/cpp of p rotated -90} ..
   endfor
   point infinity of p
         {direction infinity of p rotated 90}
 fi
enddef;
vardef zigzag expr p =
 save zpp;
 numeric zpp;
 zpp = ceiling (pixlen (p, 10) / zigzag_len) / length p;
 if not cycle p:
   point 0 of p --
 fi
 for k = 0 upto zpp*length(p) - 1:
   point (k+1/4)/zpp of p shifted
     (zigzag_width
      * dir angle (direction (k+1/4)/zpp of p rotated 90)) --
   point (k+3/4)/zpp of p shifted
     (zigzag_width
      * dir angle (direction (k+3/4)/zpp of p rotated -90)) --
 endfor
 if cycle p:
   cycle
 else:
   point infinity of p
 fi
enddef;

def midarrowperc (expr p, t) =
  fill arrowhead subpath(0,arctime(t*arclength(p)+0.5ahlength) of p) of p;
enddef;

def midarrow (expr p, t) =
  fill arrowhead subpath(0,arctime(arclength(subpath (0,t) of p)+0.5ahlength) of p) of p;
enddef;

def drawmidarrow expr p =
  draw p;
  midarrowperc(p,0.6);
enddef;

def filldot(expr z,c)=
fill fullcircle scaled dotsize shifted z withcolor c;
draw fullcircle scaled dotsize shifted z normalpen;
enddef;

pair vpos[];
path paths[];

xu:=1cm;

\end{mpostcmd}


\title{A Quantum Affine Algebra for\texorpdfstring{\\}{ }the Deformed Hubbard Chain}
\author{Niklas Beisert\texorpdfstring{$^a$}{}, Wellington Galleas\texorpdfstring{$^a$}{}\texorpdfstring{ and}{,} Takuya Matsumoto\texorpdfstring{$^{a,b}$}{}}
\subject{MSC (2010): 17B37, 17B67, 81U15; PACS (2010): 02.30.Ik, 02.20.Uw, 71.10.Fd}



\begin{document}

\pdfbookmark[1]{Title Page}{title}

\thispagestyle{empty}
\begin{flushright}\footnotesize
\texttt{\arxivlink{1102.5700}}\\
\texttt{AEI-2011-005}
\end{flushright}
\vspace{1cm}

\begin{center}%
{\Large\bfseries\thetitle\par}\vspace{1cm}

\textsc{\theauthor}\vspace{5mm}

$^a$ \textit{Max-Planck-Institut f\"ur Gravitationsphysik\\%
Albert-Einstein-Institut\\%
Am M\"uhlenberg 1, 14476 Potsdam, Germany}\vspace{3mm}%

$^b$ \textit{Graduate School of Mathematics, Nagoya University,\\
Nagoya 464-8602, Japan}\vspace{3mm}%

\verb+{nbeisert,wgalleas}@aei.mpg.de+, 
\verb+m05044c@math.nagoya-u.ac.jp+
\par\vspace{1cm}

\textbf{Abstract}\vspace{7mm}

\begin{minipage}{12.7cm}
The integrable structure of the one-dimensional Hubbard model
is based on Shastry's R-matrix and the Yangian of a centrally extended $\alg{sl}(2|2)$ superalgebra.
Alcaraz and Bariev have shown that the model admits an integrable deformation 
whose R-matrix has recently been found. This R-matrix is of trigonometric type and 
here we derive its underlying exceptional quantum affine algebra. 
We also show how the algebra reduces to the above mentioned Yangian 
and to the conventional quantum affine $\alg{sl}(2|2)$ algebra in two special limits.
\end{minipage}

\end{center}

\newpage


\section{Introduction and Overview}

The algebraic structures underlying integrable models have been intensively studied in the
past years and a variety of approaches have been formulated
in order to systematically derive solutions 
of the Yang-Baxter equation \cite{Bazhanov:1984gu,Bazh2,Galleas:2006kd,Ge,Jimbo:1985ua}. 
The solutions of the Yang-Baxter equation, also known as R-matrices, 
characterize the integrability of the model and
a large number of solutions have been obtained through the quantum group framework 
making use of deformations of universal enveloping algebras.
One of the most prominent applications of quantum groups, or more specifically quantum deformations 
$\halg{U}_{q}[\alg{g}]$ in the case considered here, lies in the fact that if $\alg{g}$ 
is finite-dimensional we can associate an operator $\mbox{R} \in \mbox{End}( \mathcal{A} \otimes \mathcal{A})$ 
satisfying the quantum Yang-Baxter equation to any representation $\mathcal{A}$ of $\halg{U}_{q}[\alg{g}]$. 
This fact was realized independently by Drinfel'd and Jimbo \cite{drinfeld,Jimbo:1985zk} 
who showed how to associate a family of Hopf algebras to any symmetrizable Kac-Moody algebra. Nevertheless, it
is worth to remark here that the defining relations of the quantum deformed algebra $\halg{U}_{q}[\alg{g}]$
first appeared in the work of Kulish and Reshetikhin on the quantum Sine-Gordon model \cite{kulish}.
The definitions of $\halg{U}_q[\alg{g}]$ can be extended 
to arbitrary Kac-Moody algebras, in particular to the affine Lie (super) algebra $\widehat{\alg{g}}$ associated with $\alg{g}$,  
and the distinction between a Lie (super) algebra and its affine extension has remarkable consequences. 

It is well known that the Yang-Baxter equation has an intimate connection with
Artin's braid group \cite{Artin} when an R-matrix does not depend
on spectral parameters \cite{Akutsu}. The constant solutions of the
Yang-Baxter equation are usually, though not always, prevenient from non-affine
Lie algebras $\alg{g}$ and the introduction of the spectral parameter can be performed in
two principal ways. The first one is the so called \emph{Baxterization} method
developed by V.\ F.\ Jones \cite{Jones}. This method makes use of the algebraic structures related to 
Artin's braid group as a starting point to derive spectral parameter dependent solutions
of the Yang-Baxter equation. The second method is based on affine Lie
algebras $\widehat{\alg{g}}$, more specifically quantum affine algebras
$\halg{U}_q[\widehat{\alg{g}}]$ or Yangian algebras $\halg{Y}[\alg{g}]$ as a special case. 
For the latter the parameter of the evaluation representation lifting the representations of $\alg{g}$ to $\widehat{\alg{g}}$ 
becomes the spectral parameter of the R-matrix.

Within the quantum group framework, the R-matrix describing scattering on the string worldsheet 
in the context of the AdS/CFT correspondence
(see \cite{Arutyunov:2009ga,Beisert:2010jr,Ahn:2010ka} for reviews)
can be obtained from a central extension of $\alg{sl}(2|2)$
\cite{Beisert:2005tm,Gomez:2006va,Plefka:2006ze}
and its Yangian algebra $\halg{Y}$ \cite{Beisert:2007ds} (see also \cite{deLeeuw:2010nd,Torrielli:2010kq}). 
Curiously, the spectral parameter dependent R-matrix in the fundamental representation
already follows from the non-affine algebra \cite{Beisert:2005tm}. 
This property however does not carry over to higher representations
where the Yangian most conveniently determines the R-matrix \cite{Arutyunov:2009pw}. 

Interestingly enough the fundamental R-matrix associated to the centrally extended $\alg{sl}(2|2)$ superalgebra
turns out to be equivalent \cite{Beisert:2006qh} to the Shastry's R-matrix \cite{shastry} responsible for
the integrable structure of the one-dimensional Hubbard model. 
The Hubbard model (see \cite{Essler:2005aa}) is the simplest
generalization beyond the band theory description of metals and it has found applications in a variety of
contexts. It can be used to describe the Mott metal-insulator transition \cite{liebwu}, $\pi$ electrons in the
benzene molecule \cite{heilmann} as well as some higher loop planar anomalous dimensions of local operators in 
$\mathcal{N}=4$ super Yang-Mills theory \cite{Rej:2005qt}. 
Now it is clear that the one-dimensional Hubbard model takes a
solitary place among the spin chain models, not just phenomenologically, but also algebraically. 
This can be observed in the Lieb-Wu equations \cite{liebwu} which have a peculiar form which is unlike
the ones for conventional spin chains based on a generic Lie (super) algebra $\alg{g}$. 
Moreover, Shastry's R-matrix is non-standard in the sense that it depends non-trivially
on \emph{two spectral parameters}, rather than a simple combination of them.
On the algebraic level these unique features can be traced to the exceptional nature of $\alg{psl}(2|2)$
which is the only simple Lie superalgebra with a non-trivial \emph{three-fold central extension} 
\cite{Nahm:1977tg,Iohara:2001aa}. Though the existence of such a large center 
allows more freedom in setting up the integrable structure,
and it is thus ultimately responsible for the peculiar features
of this model, these non-standard features have left scientists puzzled for a long time. Even now the algebraic structures
underlying the integrability of the one-dimensional Hubbard model are
far less developed than the ones for conventional spin chains, cf.\ \cite{Essler:2005aa} and \cite{Torrielli:2010kq}. 
Merely the classical limit of the algebra and its classical r-matrix is reasonably well understood \cite{Torrielli:2007mc,Beisert:2007ty}.

The one-dimensional Hubbard hamiltonian is also a paradigm in condensed matter physics, and together  with the supersymmetric
$t$-$J$ model \cite{TJ1,TJ2,Sutherland:1975vr,TJ4}, they are the fundamental blocks for the  study of non-perturbative effects 
in strongly correlated electrons systems due to the fact that they are integrable. In \cite{alcbar} Alcaraz and Bariev proposed a Bethe ansatz
solvable hamiltonian interpolating between the Hubbard and the supersymmetric $t$-$J$ models. Besides the hopping term (kinetic energy) this model contains not
only a Coulomb interaction as in the case of the Hubbard model, but also a spin-spin interaction 
resembling the $t$-$J$ hamiltonian. It turns out that this Alcaraz-Bariev model can be viewed as a quantum deformation of the Hubbard model \cite{Beisert:2008tw}
in much the same way that the Heisenberg XXZ model is a quantum deformation of the XXX model. 
More precisely, the R-matrix of the Alcaraz-Bariev model 
is based on a quantum deformation $\halg{Q}$ of the extended $\alg{sl}(2|2)$ algebra.%
\footnote{The algebra has also been discussed in the Faddeev-Zamolodchikov framework in \cite{Murgan:2008zu}.}
Though the R-matrix is not necessary in order to obtain the exact spectrum of the model, this knowledge still offers the possibility of studying
thermodynamic properties in an efficient way through the quantum transfer matrix method \cite{qtm}.

Much of the same peculiar features of the Hubbard model applies to the Alcaraz-Bariev model and the associated
quantum deformation $\halg{Q}$ of the centrally extended $\alg{sl}(2|2)$ algebra \cite{Beisert:2008tw}.
However, with the caveat that quantum deformation makes some structures substantially more complicated to handle.
Except for its classical limit \cite{Beisert:2010kk}, which already provides valuable insights into the expected structures,
it is fair to say that our knowledge of the complete underlying algebra is still limited. With that in mind,
the scenario described above thus asks for a formulation of the \emph{quantum affine algebra} $\widehat{\halg{Q}}$
based on the extended $\alg{sl}(2|2)$. Even though quantum deformations introduce additional complexity,
they also bring about some new symmetries into the framework as compared to Yangians which are rather singular limits thereof.
This may eventually help us to uncover the full structure of the Hopf algebra underlying integrability in the AdS/CFT correspondence. 

\medskip

This paper is organized as follows:
We start in \secref{sec:nonaffine} with a review of the 
quantum deformed extended $\alg{sl}(2|2)$ algebra $\halg{Q}$
and its associated integrable structures.
Next we use a special property of its affine Dynkin diagram
to derive the affine extension $\widehat{\halg{Q}}_0$
in \secref{sec:affder}. For the reader's convenience we summarize
the algebraic relations of $\widehat{\halg{Q}}_0$ in \secref{sec:summary}.
We go on by establishing the fundamental representation in 
\secref{sec:fundamental} which requires to refine 
the algebra $\widehat{\halg{Q}}_0$ to $\widehat{\halg{Q}}$.
In the remainder of the paper we study two interesting limits
of the algebra. One of them is the conventional quantum affine algebra
$\halg{U}_q[\widehat{\alg{sl}}(2|2)]$ described in the \secref{sec:convlimit}, followed by
the extended $\alg{sl}(2|2)$ Yangian $\halg{Y}$ discussed in the \secref{sec:yangian}.
The \secref{sec:concl} is left for conclusions and final remarks.

\section{Quantum Deformation of Extended \texorpdfstring{$\alg{sl}(2|2)$}{sl(2|2)}}
\label{sec:nonaffine}

In the following we shall briefly review the quantum deformed extended $\alg{sl}(2|2)$ algebra 
$\halg{Q}$ introduced in \cite{Beisert:2008tw}.

\paragraph{Cartan Matrix.}

\begin{mpostfile}{FigDynkinOXO.mps}
paths[1]:=fullcircle scaled 0.6xu shifted (-1.5xu,0);
paths[2]:=fullcircle scaled 0.6xu shifted (0xu,0);
paths[3]:=fullcircle scaled 0.6xu shifted (+1.5xu,0);

draw paths[1] fatpen;
draw paths[2] fatpen;
draw paths[3] fatpen;

draw (point 0 of paths[1])--(point 4 of paths[2]) fatpen;
draw (point 0 of paths[2])--(point 4 of paths[3]) fatpen;

draw (point 1 of paths[2])--(point 5 of paths[2]) fatpen;
draw (point 3 of paths[2])--(point 7 of paths[2]) fatpen;

label.bot(btex $1$ etex, point 6 of paths[1]);
label.bot(btex $2$ etex, point 6 of paths[2]);
label.bot(btex $3$ etex, point 6 of paths[3]);

putcopyspace;
putcopy;
\end{mpostfile}

We shall consider the $\alg{sl}(2|2)$ Dynkin diagram in \figref{fig:OXO}
\begin{figure}\centering
\includegraphicsbox{FigDynkinOXO.mps}
\caption{Dynkin diagram for $\alg{sl}(2|2)$.}
\label{fig:OXO}
\end{figure}
such that the associated Cartan matrix $A$ and normalization matrix $D$ read
\[
A=\matr{rrr}{+2&-1&0\\+1&0&-1\\0&-1&+2}
\qquad
D=\mbox{diag}(+1,-1,-1).
\]
With the help of $D$, we obtain the following symmetric matrix
which frequently appears in the defining relations
\[
DA=\matr{rrr}{+2&-1&0\\-1&0&+1\\0&+1&-2}.
\]

\paragraph{Generators.}

The algebra is conveniently presented in terms of Chevalley-Serre generators.
The generators are the raising and lowering generators 
$\gen{E}_j$ and $\gen{F}_j$
as well as the exponentiated Cartan generators 
$\gen{K}_j=q^{\gen{H}_j}$ with $j=1,2,3$. 
All of them are even generators of our superalgebra, 
except for the pair of odd generators $\gen{E}_2$ and $\gen{F}_2$, 
in accordance with the Dynkin diagram in \figref{fig:OXO}.
In addition there are \emph{two} central charges $\gen{U}$ and $\gen{V}=q^{\gen{C}}$.
The algebra has two parameters: the deformation parameter $q$ and
the coupling parameter $g$. A third parameter $\alpha$ could be 
absorbed into a redefinition of the generators,
and thus does not count as a parameter of the algebra. 
Nevertheless it is convenient to keep it unspecified.

\paragraph{Algebra.}

The Chevalley-Serre generators satisfy the standard quantum deformed commutation 
relations 
($j,k=1,2,3$)%
\footnote{Our $\gen{K}_2$, $\gen{K}_3$ are inverted compared to usual conventions
to make the symmetric matrix $DA$ appear in place of the Cartan matrix $A$;
this makes $D_{jj}$ appear in $\scomm{\gen{E}_j}{\gen{F}_k}$.}
\[
\label{algebra}
\gen{K}_j\gen{E}_k=q^{DA_{jk}}\gen{E}_k\gen{K}_j
\quad \quad 
\gen{F}_k\gen{K}_j=q^{DA_{jk}}\gen{K}_j\gen{F}_k
\quad \quad
\scomm{\gen{E}_j}{\gen{F}_k}=D_{jj}\delta_{jk}
\frac{\gen{K}_j-\gen{K}_j^{-1}}{q-q^{-1}} \; .
\]
In addition, the following Serre relations hold
($j=1,3$)
\begin{align}
\comm{\gen{E}_1}{\gen{E}_3}
=\acomm{\gen{E}_2}{\gen{E}_2}
=\bigcomm{\gen{E}_j}{\comm{\gen{E}_j}{\gen{E}_2}}
-(q-2+q^{-1})\gen{E}_j\gen{E}_2\gen{E}_j&=0
\nln
\comm{\gen{F}_1}{\gen{F}_3}
=\acomm{\gen{F}_2}{\gen{F}_2}
=\bigcomm{\gen{F}_j}{\comm{\gen{F}_j}{\gen{F}_2}}
-(q-2+q^{-1})\gen{F}_j\gen{F}_2\gen{F}_j&=0 \; .
\end{align}
%

\paragraph{Center.}

The algebra defined by the above relations has three central elements
\begin{align}\label{eq:center1}
\gen{C}_1&=\gen{K}_1\gen{K}_2^2\gen{K}_3 
\nln
\gen{C}_2&=\bigacomm{\comm{\gen{E}_2}{\gen{E}_1}}{\comm{\gen{E}_2}{\gen{E}_3}}
-(q-2+q^{-1})\gen{E}_2\gen{E}_1\gen{E}_3\gen{E}_2 
\nln
\gen{C}_3&=\bigacomm{\comm{\gen{F}_2}{\gen{F}_1}}{\comm{\gen{F}_2}{\gen{F}_3}}
-(q-2+q^{-1})\gen{F}_2\gen{F}_1\gen{F}_3\gen{F}_2 \; .
\end{align}
The latter two are usually projected out by the Serre relations 
$\gen{C}_2=\gen{C}_3=0$ of the superalgebra $\alg{sl}(2|2)$.
Furthermore, in $\alg{psl}(2|2)$ also the former is projected out by the condition $\gen{C}_1=1$.
Here we keep them all and thus our algebra is based on a central extension 
of $\alg{psl}(2|2)$ or $\alg{sl}(2|2)$.
As shown in \cite{Plefka:2006ze,Beisert:2008tw}, 
it turns out that we obtain a very interesting algebra if 
we impose one constraint on the central elements as follows:
\[\label{eq:center2}
\gen{C}_1=\gen{V}^{-2}
\qquad
\gen{C}_2=g\alpha(1-\gen{U}^2\gen{V}^2)
\qquad
\gen{C}_3=g\alpha^{-1}(\gen{V}^{-2}-\gen{U}^{-2}) \; .
\]
%

\paragraph{Coalgebra.}

All the above relations are compatible with the 
following coalgebra structure.
The coproduct for all $\gen{X}\in\set{\gen{K}_j, \gen{U},\gen{V}}$ 
is group-like, $\copro(\gen{X})=\gen{X}\otimes\gen{X}$,
while for $\gen{E}_j$ and $\gen{F}_j$ it takes the standard form
but with a twist induced by the central element $\gen{U}$,
\[
\copro(\gen{E}_j)=\gen{E}_j\otimes 1+\gen{K}_j^{-1}\gen{U}^{+\delta_{j,2}}\otimes\gen{E}_j
\qquad
\copro(\gen{F}_j)=\gen{F}_j\otimes \gen{K}_j+\gen{U}^{-\delta_{j,2}}\otimes\gen{F}_j.
\]
The twist is based on the $\alg{gl}(1)$ derivation in $\alg{gl}(2|2)$ 
which applies only to the fermionic generators $\gen{E}_2$ and $\gen{F}_2$.

\paragraph{Fundamental Representation.}

The algebra has a family of representations
acting on the $(2|2)$-dimensional graded space $\mathbb{V}$.
The raising and lowering generators are represented by 
the following $(2|2)\times(2|2)$ supermatrices 
\begin{align}\label{FundReps}
\gen{E}_1&\simeq
\tinymatr{cc|cc}{0&0&0&0\\1&0&0&0\\\hline 0&0&0&0\\0&0&0&0}
&
\gen{E}_2&\simeq
\tinymatr{cc|cc}{0&0&0&b\\0&0&0&0\\\hline 0&a&0&0\\0&0&0&0}
&
\gen{E}_3&\simeq
\tinymatr{cc|cc}{0&0&0&0\\0&0&0&0\\\hline 0&0&0&0\\0&0&1&0}
\nln
\gen{F}_1&\simeq
\tinymatr{cc|cc}{0&1&0&0\\0&0&0&0\\\hline 0&0&0&0\\0&0&0&0}
&
\gen{F}_2&\simeq
\tinymatr{cc|cc}{0&0&0&0\\0&0&d&0\\\hline 0&0&0&0\\c&0&0&0}
&
\gen{F}_3&\simeq
\tinymatr{cc|cc}{0&0&0&0\\0&0&0&0\\\hline 0&0&0&1\\0&0&0&0}.
\end{align}
We shall not present here the supermatrix representations for $\gen{K}_j$ since they easily follow from
the algebra relations (\ref{algebra}). 
The central elements $\gen{U}$ and $\gen{V}$ are represented 
by uniform multiplication with $U$ and $V$ respectively. 
In their turn these central elements are related to the coefficients $a$, $b$, $c$ and $d$ through the constraints
\begin{align}
\label{FundCond}
ad&=\frac{q^{1/2}V-q^{-1/2}V^{-1}}{q-q^{-1}}
&
bc&=\frac{q^{-1/2}V-q^{1/2}V^{-1}}{q-q^{-1}}
\nln
ab&= g\alpha(1-U^2V^2)
&
cd&= g\alpha^{-1}(V^{-2}-U^{-2}).
\end{align}
The above constraints imply the following relation between $U$ and $V$,
\[\label{FundCond2}
g^2(V^{-2}-U^{-2})(1-U^2V^2)=\frac{(V-qV^{-1})(V-q^{-1}V^{-1})}{(q-q^{-1})^2} \; ,
\]
while one of the parameters $a,b,c,d$ can be chosen freely.
Altogether we thus have a two-parameter family of representations.

\paragraph{Fundamental R-matrix.}

In \cite{Beisert:2008tw} the fundamental R-matrix
for the above described algebra has been explicitly derived.
The R-matrix is a linear map $\rmat:\mathbb{V}\otimes \mathbb{V}\to \mathbb{V}\otimes \mathbb{V}$
which is a function of the variables parametrizing each one of the spaces $\mathbb{V}$.
The form of the R-matrix was obtained by demanding that the cocommutativity condition
\[
\label{coco}
\rmat \copro(\gen{X}) = \coproop(\gen{X}) \rmat 
\]
holds for $\gen{X} \in \set{\gen{E}_j,\gen{F}_j,\gen{K}_j,\gen{U},\gen{V}}$.
Here $\coproop(\gen{X})$ stands for the opposite coproduct defined through the
permutation map
\[
\coproop(\gen{X}) = \mathcal{P} \copro(\gen{X})  \mathcal{P} \;
\]
where $\mathcal{P}$ denotes the graded permutation operator. 
The relation \eqref{coco} has proved to completely and consistently determine
the fundamental R-matrix up to an overall scalar factor.
The explicit form of $\rmat$ is lengthy and shall not be reproduced here since 
it was given in \cite{Beisert:2008tw}.

\section{Derivation of the Affine Extension}
\label{sec:affder}

\begin{mpostfile}{FigDynkinOXOX.mps}
paths[1]:=fullcircle scaled 0.6xu shifted (-1xu,0);
paths[2]:=fullcircle scaled 0.6xu shifted (0xu,-1xu);
paths[3]:=fullcircle scaled 0.6xu shifted (+1xu,0);
paths[4]:=fullcircle scaled 0.6xu shifted (0xu,+1xu);

draw paths[1] fatpen;
draw paths[2] fatpen;
draw paths[3] fatpen;
draw paths[4] fatpen;

draw (point 7 of paths[1])--(point 3 of paths[2]) fatpen;
draw (point 1 of paths[2])--(point 5 of paths[3]) fatpen;

draw (point 1 of paths[1])--(point 5 of paths[4]) fatpen;
draw (point 7 of paths[4])--(point 3 of paths[3]) fatpen;

draw (point 1 of paths[2])--(point 5 of paths[2]) fatpen;
draw (point 3 of paths[2])--(point 7 of paths[2]) fatpen;

draw (point 1 of paths[4])--(point 5 of paths[4]) fatpen;
draw (point 3 of paths[4])--(point 7 of paths[4]) fatpen;

label.lft(btex $1$ etex, point 4 of paths[1]);
label.bot(btex $2$ etex, point 6 of paths[2]);
label.rt(btex $3$ etex, point 0 of paths[3]);
label.top(btex $4$ etex, point 2 of paths[4]);

putcopyspace;
putcopy;
\end{mpostfile}

Now we shall consider the affine extension of the algebra defined above. 
The affine extension for the Dynkin diagram in \figref{fig:OXO} 
is given in \figref{fig:OXOX}.
\begin{figure}\centering
\includegraphicsbox{FigDynkinOXOX.mps}
\caption{Dynkin diagram for $\widehat{\alg{sl}}(2|2)$.}
\label{fig:OXOX}
\end{figure}
The associated Cartan matrix $A$ for $\widehat{\alg{sl}}(2|2)$
and the symmetric Cartan matrix $DA$ with 
$D=\mbox{diag}(+1,-1,-1,-1)$
now read
\[
A=\matr{rrrr}{+2&-1&0&-1\\+1&0&-1&0\\0&-1&+2&-1\\+1&0&-1&0}
\qquad
DA=\matr{rrrr}{+2&-1&0&-1\\-1&0&+1&0\\0&+1&-2&+1\\-1&0&+1&0}\;.
\]
The crucial observation here is that the new
fourth node of the Dynkin diagram is completely analogous to the second one. 
Consequently the second and fourth rows and columns of the matrix $DA$ coincide.
In practice that means that the associated Chevalley-Serre generators should
obey analogous commutation relations. This observation will help us tremendously in 
completing this unusual affine algebra.

\paragraph{Doubling the Fermionic Node.}

We introduce the new set of generators $\{ \gen{E}_4,\gen{F}_4,\gen{K}_4 \}$ and, as explained above, they
should act as copies of the generators $\{ \gen{E}_2,\gen{F}_2,\gen{K}_2 \}$.
In their turn the coupling constant $g$, the normalization $\alpha$ as well as the central elements $\gen{U}$ and $\gen{V}$ 
always appear in conjunction with the generators $ \{ \gen{E}_2,\gen{F}_2,\gen{K}_2 \}$. 
Thus it makes sense to double those as well in such a way that we relabel 
$\{ g, \alpha,\gen{U},\gen{V} \}$ as $\{ g_2, \alpha_2, \gen{U}_2, \gen{V}_2 \}$,
and introduce new constants and central generators $\{ g_4,\alpha_4,\gen{U}_4,\gen{V}_4 \}$.

The algebra relations and coproducts for the new generators 
$ \{ \gen{E}_4,\gen{F}_4,\gen{K}_4 \}$ will be direct copies of the ones for
$ \{ \gen{E}_2,\gen{F}_2,\gen{K}_2 \}$ discussed in \secref{sec:nonaffine}. This almost guarantees that we get a consistent algebra and
coalgebra structure. Now we merely have to take care of the relations of 
the quantum affine algebra $\widehat{\alg{sl}}(2|2)$ mixing the two sets of generators, namely the anticommutators
$\acomm{\gen{E}_2}{\gen{F}_4}$, $\acomm{\gen{E}_4}{\gen{F}_2}$,
$\acomm{\gen{E}_2}{\gen{E}_4}$ and $\acomm{\gen{F}_2}{\gen{F}_4}$.

\paragraph{Compatibility.}

The anticommutators $\acomm{\gen{E}_2}{\gen{F}_4}$ and $\acomm{\gen{E}_4}{\gen{F}_2}$ 
commute with the Cartan subalgebra and thus they should belong to it as well.
Fortunately the coproducts for the generators involved are completely fixed at this stage
and the compatibility between them imposes constraints over the algebra.
In particular we have 
\[
\copro(\gen{E}_2)=\gen{E}_2\otimes 1+\gen{K}_2^{-1}\gen{U}_2\otimes\gen{E}_2
\qquad
\copro(\gen{F}_4)=\gen{F}_4\otimes \gen{K}_4+\gen{U}_4^{-1}\otimes\gen{F}_4,
\]
and thus
\[
\acomm{\copro(\gen{E}_2)}{\copro(\gen{F}_4)}
=
\acomm{\gen{E}_2}{\gen{F}_4}\otimes \gen{K}_4
+\gen{K}_2^{-1}\gen{U}_2\gen{U}_4^{-1}\otimes\acomm{\gen{E}_2}{\gen{F}_4}.
\]
This suggests that $\acomm{\gen{E}_2}{\gen{F}_4}$ should be composed by a linear combination 
of the group-like elements $\gen{K}_4$ and $\gen{K}_2^{-1}\gen{U}_2\gen{U}_4^{-1}$. Under these
considerations we can use an ansatz and easily obtain a solution for the compatibility condition
$\acomm{\copro(\gen{E}_2)}{\copro(\gen{F}_4)}=\copro(\acomm{\gen{E}_2}{\gen{F}_4})$. By doing so we find
\begin{equation}
\label{mix1}
\acomm{\gen{E}_2}{\gen{F}_4}= -\tilde g\tilde\alpha^{-1}(\gen{K}_4-\gen{U}_4^{-1}\gen{U}^{}_2\gen{K}_2^{-1}) 
\end{equation}
and similarly
\begin{equation}
\label{mix2}
\acomm{\gen{E}_4}{\gen{F}_2}= +\tilde g\tilde\alpha^{+1}(\gen{K}_2-\gen{U}_2^{-1}\gen{U}^{}_4\gen{K}_4^{-1})
\end{equation}
with two new constants $\tilde g$ and $\tilde\alpha$. In the standard quantum affine algebra $\widehat{\alg{sl}}(2|2)$
the r.h.s.\  of (\ref{mix1}) and (\ref{mix2}) vanishes and this is one of the main differences of our 
unusual affine algebra. It is worth to remark here that similar relations, though not 
equivalent, also appeared in \cite{Yamane}.

The anticommutators $\acomm{\gen{E}_2}{\gen{E}_4}$ and $\acomm{\gen{F}_2}{\gen{F}_4}$
do not commute with the Cartan subalgebra and considerations on the coalgebra structure lead us to
conclude that they must be trivial. Hence
\[
\acomm{\gen{E}_2}{\gen{E}_4}=\acomm{\gen{F}_2}{\gen{F}_4} = 0 .
\]

The question remains whether the above relations,
in particular the mixed ones \eqref{mix1,mix2},
define a consistent algebra:
As we shall see later, the algebra admits at least one representation. 
Using the coproduct, one can define further representations as 
tensor products. Hence the relations consistently define 
an algebra with a non-trivial representation theory.%
\footnote{It is conceivable though that the above relations 
imply further simple relations,
such as $\gen{U}_2\gen{U}_4=1$ and $\gen{V}_2\gen{V}_4=1$ which
hold on the representation in \secref{sec:fundamental}.}

\section{Hopf Algebra Structure}
\label{sec:summary}

We shall call the above derived quantum affine algebra $\widehat{\halg{Q}}_0$ and
in what follows we summarize its defining relations.
Some of the constants will be refined later to give a more special algebra $\widehat{\halg{Q}}$.

\paragraph{Algebra.}

The algebra $\widehat{\halg{Q}}_0$ consists of a deformed extension of the quantum affine algebra 
$\widehat{\alg{sl}}(2|2)$.
It is generated by the corresponding Chevalley-Serre generators 
$\gen{K}_j,\gen{E}_j,\gen{F}_j$ ($i,j=1,2,3,4$)
and central elements $\gen{U}_k$ and $\gen{V}_k$ ($k=2,4$).
It is also useful to recall here the symmetric matrix $DA$ 
and the normalization matrix $D$ associated to the Cartan matrix $A$ for $\widehat{\alg{sl}}(2|2)$:
\[
\label{cmat}
DA=\matr{rrrr}{+2&-1&0&-1\\-1&0&+1&0\\0&+1&-2&+1\\-1&0&+1&0} 
\qquad
D=\mbox{diag}(+1,-1,-1,-1).
\]

The algebra has a set of group-like elements
$\gen{X},\gen{Y}\in\set{1,\gen{K}_j,\gen{U}_k,\gen{V}_k}$
which are invertible and commutative
\[
\gen{X}\gen{X}^{-1}=1
\qquad \qquad
\gen{X}\gen{Y}=\gen{Y}\gen{X}.
\]

The Chevalley-Serre raising and lowering generators $\gen{E}_j$ and $\gen{F}_j$ satisfy 
the usual relations, except for the two mixed anticommutators given in (\ref{mix1}) and (\ref{mix2}),
\begin{align}
\label{psl22}
\gen{K}_i \gen{E}_j \gen{K}^{-1}_i  &= q^{DA_{ij}} \gen{E}_j \qquad
&
\gen{K}_i\gen{F}_j \gen{K}^{-1}_i  &=q^{-DA_{ij}} \gen{F}_j 
\nln
\acomm{\gen{E}_2}{\gen{F}_4}&= -\tilde g \tilde\alpha^{-1} \bigbrk{\gen{K}_4-\gen{U}_2 \gen{U}_{4}^{-1}  \gen{K}_2^{-1}} \qquad
&
\acomm{\gen{E}_4}{\gen{F}_2}&=  \tilde g \tilde\alpha\bigbrk{\gen{K}_2 - \gen{U}_4 \gen{U}_{2}^{-1} \gen{K}_4^{-1}}
\nln
\scomm{\gen{E}_j}{\gen{F}_j} &= D_{jj}\frac{\gen{K}_j-\gen{K}_j^{-1}}{q-q^{-1}} \qquad
&
\scomm{\gen{E}_i}{\gen{F}_j}&=0 \quad \mbox{for $i\neq j$, $i+j\neq 6$} \; .
\end{align}
In addition to the relations (\ref{psl22}), the algebra $\widehat{\halg{Q}}_0$ also 
satisfy the following Serre relations ($j=1,3$),
\begin{align}
\label{serre}
\comm{\gen{E}_1}{\gen{E}_3} = \gen{E}_2 \gen{E}_2 = \gen{E}_4 \gen{E}_4 
=\acomm{\gen{E}_2}{\gen{E}_4}
&= 0
\nln
\comm{\gen{F}_1}{\gen{F}_3} = \gen{F}_2 \gen{F}_2 = \gen{F}_4 \gen{F}_4
=\acomm{\gen{F}_2}{\gen{F}_4}
 &= 0
\nln
\bigcomm{\gen{E}_j}{\comm{\gen{E}_j}{\gen{E}_k}} - (q-2+q^{-1})\gen{E}_j\gen{E}_k\gen{E}_j &= 0 
\nln
\bigcomm{\gen{F}_j}{\comm{\gen{F}_j}{\gen{F}_k}} - (q-2+q^{-1})\gen{F}_j\gen{F}_k\gen{F}_j &= 0 \; .
\end{align}
The quartic Serre relations of the superalgebra $\widehat{\alg{sl}}(2|2)$ 
are deformed by the central elements $\gen{U}_k$ and $\gen{V}_k$ 
as follows,
\begin{align}
\label{qserre}
\bigacomm{\comm{\gen{E}_1}{\gen{E}_k}}{\comm{\gen{E}_3}{\gen{E}_k}}
-(q-2+q^{-1})\gen{E}_k\gen{E}_1\gen{E}_3\gen{E}_k &= g_k\alpha_k(1-\gen{V}_k^2\gen{U}_k^2)
\nln
\bigacomm{\comm{\gen{F}_1}{\gen{F}_k}}{\comm{\gen{F}_3}{\gen{F}_k}}
-(q-2+q^{-1})\gen{F}_k\gen{F}_1\gen{F}_3\gen{F}_k &= g_k\alpha^{-1}_k(\gen{V}_k^{-2}-\gen{U}_k^{-2}) \; ,
\end{align}
and the remaining central elements of the superalgebra $\widehat{\alg{sl}}(2|2)$ 
are then related to $\gen{V}_k$ through
\[
\label{Kk}
\gen{K}^{-1}_{1} \gen{K}^{-2}_{k} \gen{K}^{-1}_{3}=\gen{V}_k^2.
\]
In summary, the above quantum affine algebra $\widehat{\halg{Q}}_0$ 
has five parameters:  $q$, $g_k$, $\tilde g$ and $\tilde\alpha$.
The two normalizations $\alpha_k$ 
merely originate from our choice of basis.

\paragraph{Algebra Automorphism.}

The quantum affine algebra $\widehat{\halg{Q}}_0$ has been constructed by making use of the similarity between
the nodes $2$ and $4$ of the Dynkin diagram in \figref{fig:OXOX}.
In fact this similarity leads to an algebra automorphism
flipping the nodes $2$ and $4$
if the coupling constants are related by
\[
\label{eq:autopar}
g_2= g_4
\qquad \qquad
\alpha_4= \zeta^2\tilde\alpha^2 \alpha_2
\]
where $\zeta^4 = 1$. Thus the following map is an algebra automorphism
\begin{align}\label{eq:auto}
\gen{E}_2&\to\zeta\tilde\alpha^{-1} \gen{E}_4
&
\gen{E}_4&\to-\zeta\tilde\alpha \gen{E}_2
\nln
\gen{F}_2&\to \zeta^{-1}\tilde\alpha \gen{F}_4
&
\gen{F}_4&\to-\zeta^{-1}\tilde\alpha^{-1} \gen{F}_2
\nln
\gen{U}_2&\to\gen{U}_4
&
\gen{U}_4&\to\gen{U}_2
\nln
\gen{K}_2&\to\gen{K}_4
&
\gen{K}_4&\to\gen{K}_2 \; .
\end{align}
%

\paragraph{Coalgebra, Antipode and Counit.}

For the group-like elements $\gen{X}\in\set{1,\gen{K}_j,\gen{U}_k,\gen{V}_k}$ ($j=1,2,3,4$ and $k=2,4$)
the coproduct $\copro$, the antipode $\antipode$ and the counit $\counit$ are defined as usual,
\[
\copro(\gen{X})=\gen{X}\otimes \gen{X} 
\qquad \qquad
\antipode(\gen{X})=\gen{X}^{-1}
\qquad \qquad
\counit(\gen{X})=1 \; ,
\]
while for the remaining Chevalley-Serre generators they are deformed by the central elements $\gen{U}_k$ 
as follows ($j=1,2,3,4$),
\begin{align}\label{coprod}
\copro(\gen{E}_j)&=\gen{E}_j\otimes 1+\gen{K}_j^{-1} \gen{U}_{2}^{+\delta_{j,2}} \gen{U}_{4}^{+\delta_{j,4}}  \otimes\gen{E}_j
\quad &
\antipode(\gen{E}_j)&=-\gen{U}_{2}^{-\delta_{j,2}} \gen{U}_{4}^{-\delta_{j,4}}\gen{K}_j \gen{E}_j  
\quad&
\counit(\gen{E}_j)&= 0
\nln
\copro(\gen{F}_j)&=\gen{F}_j\otimes \gen{K}_j+ \gen{U}_{2}^{-\delta_{j,2}} \gen{U}_{4}^{-\delta_{j,4}} \otimes\gen{F}_j 
&
\antipode(\gen{F}_j)&=-\gen{U}_{2}^{+\delta_{j,2}} \gen{U}_{4}^{+\delta_{j,4}} \gen{F}_j\gen{K}_j^{-1}
&
\counit(\gen{F}_j) &= 0 .
\end{align}
The above relations characterize our quantum affine algebra $\widehat{\halg{Q}}_0$ as a Hopf algebra.
We have verified explicitly the compatibility between the algebra and 
the coalgebra (as well as the antipode relations). 
In other words, $\copro(XY)=\copro(X)\copro(Y)$ is compatible with 
all algebra relations. 
In particular, the unusual algebra relations 
\eqref{eq:center1,eq:center2,mix1,mix2} were derived
in order to obtain a consistent Hopf algebra structure.
In the following section we shall discuss the algebra's fundamental representation,
upon which a large class of finite-dimensional representations can
be constructed by means of the coalgebra.

\section{Fundamental Representation}
\label{sec:fundamental}

Now we would like to lift the 4-dimensional fundamental representation given 
in (\ref{FundReps}) to a representation of the affine algebra.
The representation theory of affine algebras has been discussed in \cite{chari_pressley1,chari_pressley1a}. In particular,
it was shown in \cite{chari_pressley2} that any finite-dimensional irreducible representation  of
$\widehat{\alg{g}}$ extended from $\alg{g}$ is isomorphic to an evaluation representation. In the quantum case there
also exist an evaluation homomorphism $ev: \halg{U}_{q}[\alg{g}] \rightarrow \halg{U}_{q}[\widehat{\alg{g}}]$ defined by Jimbo in 
\cite{Jimbo:1985vd,Jimbo:1985ua} which reduces to the usual evaluation in the classical limit $q \rightarrow 1$. Moreover, when 
$\alg{g} \cong  \alg{sl}$ it was shown in
\cite{chari_pressley} that any extension of a representation from $\halg{U}_{q}[\alg{g}]$ to $\halg{U}_{q}[\widehat{\alg{g}}]$
on the same space is isomorphic to an evaluation representation. 

Due to the non-standard nature of our extended quantum affine algebra $\widehat{\halg{Q}}_0$, 
it is not clear if this whole scenario of evaluation representations applies to our case. 
Nevertheless we find here that the set of generators 
$\{ \gen{K}_4 , \gen{E}_4 , \gen{F}_4 , \gen{U}_4 , \gen{V}_4 \}$ 
satisfying (\ref{psl22}) can be obtained
as copies of the generators 
$\{ \gen{K}_2 , \gen{E}_2 , \gen{F}_2 , \gen{U}_2 , \gen{V}_2 \}$ 
with modified coefficients.

\paragraph{Doubling Ansatz.}
As before we assume the generators $ \gen{E}_4$ and $\gen{F}_4 $
to act respectively as copies of $ \gen{E}_2$ and $\gen{F}_2$ but with different coefficients. 
Hence,
\[
\label{FundReps2}
\gen{E}_k\simeq
\tinymatr{cc|cc}{0&0&0&b_k\\0&0&0&0\\\hline 0&a_k&0&0\\0&0&0&0}
\qquad \mbox{and} \qquad
\gen{F}_k\simeq
\tinymatr{cc|cc}{0&0&0&0\\0&0&d_k&0\\\hline 0&0&0&0\\c_k&0&0&0} \qquad \mbox{for} \;\;\; k=2,4 .
\]
By doing so we obtain two sets of four constraints from (\ref{FundCond}).
Furthermore, the mixed relations (\ref{mix1}) and (\ref{mix2}) yield another set of four constraints, 
namely
\begin{align}
\label{FundCond3}
a_2d_4&= \tilde g\tilde\alpha^{-1}\bigbrk{q^{1/2}U_2U_4^{-1}V_2-q^{-1/2}V_4^{-1}}
&
b_2c_4&= \tilde g\tilde\alpha^{-1}\bigbrk{q^{-1/2}U_2U_4^{-1}V_2-q^{1/2}V_4^{-1}}
\nln
c_2b_4&= \tilde g\tilde\alpha\bigbrk{q^{1/2}V_2^{-1}-q^{-1/2}U_2^{-1}U_4 V_4}
&
d_2a_4&= \tilde g\tilde\alpha\bigbrk{q^{-1/2}V_2^{-1}-q^{1/2}U_2^{-1}U_4 V_4 } \;\; .
\end{align}
In total we have 12 constraints for 12 parameters $(a_k,b_k,c_k,d_k,U_k,V_k)$.
Thus the solution of the constraints completely fixes all the parameters and leaves
just a discrete set of 4-dimensional representations.

\paragraph{Constrained Parameters.}

The seven constants $g_k,\alpha_k,\tilde g,\tilde \alpha,q$ can be chosen in a special way
in order to solve two of the constraints.
One suitable choice%
\footnote{Another choice that will not be discussed here is
$\tilde g^2=-1/(q-q^{-1})^2$ and 
$\alpha_4=-\alpha_2\tilde\alpha^2(g_2/g_4)^{\pm 1}$.}
expressed in terms of the four parameters $g,q,\alpha,\tilde\alpha$ reads%
\footnote{It would be interesting to see what implications these relations 
might have on the algebra relations defined in \secref{sec:summary}
as they change the representation theory substantially.}
\[\label{ParRel}
g_2=g_4=g
\qquad
\alpha_2=\alpha_4\tilde\alpha^{-2}=\alpha
\qquad
\tilde g^2=\frac{g^2}{1-g^2(q-q^{-1})^2} \; .
\]
In fact there is a convenient replacement for $g$ in terms
of a new parameter $\tilde q$ 
which also allow us to parametrize the quadratic relation for $\tilde g$ as
\[\label{eq:ParRel2}
g= \frac{\tilde q-\tilde q^{-1}}{2i(q-q^{-1})}
\qquad 
\tilde g= \frac{i(\tilde q-\tilde q^{-1})}{(q-q^{-1})(\tilde q+\tilde q^{-1})} \; .
\]
We shall be mainly concerned with the above choice of parameters in this paper.
Thus we shall denote the algebra $\widehat{\halg{Q}}_0$ obeying the constraints 
\eqref{ParRel}
by $\widehat{\halg{Q}}_{g,q}$ or $\widehat{\halg{Q}}$ for short.
It depends on two parameters, $g$ and $q$, and it is expressed using two 
normalization constants $\alpha$ and $\tilde\alpha$. 
Nevertheless we shall also use the original parameters
$g_k,\alpha_k,\tilde g,\tilde \alpha,q$
with the above relations implied.

\paragraph{Two-Parameter Family.}

The solution of the remaining constraints for the fundamental representation leave us with 
\[\label{inv}
U_4=\pm U_2^{-1}
\qquad \qquad \qquad
V_4=\pm V_2^{-1}.
\]
The relations (\ref{FundCond2}) between the $U_k$ and the $V_k$
then automatically coincide.
Furthermore, one of the coefficients $a_k,b_k,c_k,d_k$ can be chosen freely.
Altogether this amounts to a two-parameter family of representations
which is thus a unique lift of the fundamental representation 
to the quantum affine algebra.

It is interesting to observe here that the representations of $\gen{E}_2$ and $\gen{F}_2$
are respectively related to the representations of $\gen{E}_4$ and $\gen{F}_4$ by the simple map given in \eqref{inv}.
In fact, this map also appears when considering the transpose representation.

\paragraph{The \texorpdfstring{$x^\pm$}{x+-}-Parametrization.}

Above we have obtained constraints for the coefficients 
$a_k , b_k, c_k , d_k$ ($k=2,4$) characterizing the fundamental representation of the
quantum affine algebra $\widehat{\halg{Q}}$. In particular, instead of solving the 
constraints (\ref{FundCond}) in favor of $U_k$ and $V_k$, as it was done in (\ref{FundCond2}),
we could also have solved them in favor of the coefficients $a_k , b_k, c_k , d_k$. 
In that case we would be left with the relation $(k=2,4)$
\[
\label{constr}
(a_k d_k - q b_k c_k) (a_k d_k - q^{-1} b_k c_k) = 1  \;.
\]

A convenient novel parametrization of this constraint 
uses a pair of variables $x^+$ and $x^-$
related by $q^{-1}\zeta(x^{+}) = q\zeta(x^{-})$ with 
\[
\label{zz}
\zeta (x) =-\frac{x + 1/x + \xi + 1/\xi}{\xi-1/\xi}
\qquad
\xi = -i \tilde g (q - q^{-1})
\; .
\]
Note that in order to simplify our results we are considering a convention for $x^{\pm}$ different from  
the one used in \cite{Beisert:2008tw}. 
More precisely, the convention used here can be obtained from the one of \cite{Beisert:2008tw}
by performing the transformation $x^{\pm}\indup{BK} = g\tilde g^{-1} (  x^{\pm}\indup{here} + \xi )$.%
\footnote{Gladly, the R-matrix in \cite{Beisert:2008tw} 
is only mildly affected by this affine transformation:
$A,D,G,H,K,L$ do not change; in $B,E$ substitute $s(x)=1/x$; only $C,F$ require more care.}

The $a_k , b_k, c_k , d_k$ can now be parametrized
in terms of variables $x_k^{\pm}$ and $\gamma_k$ as follows
\begin{align}
\label{abcd}
a_k &= \sqrt{g} \gamma_k 
&
b_k &= \frac{\sqrt{g} \alpha_k}{\gamma_k}\, \frac{x_k^{-}-x_k^{+}}{x_k^{-}}
\nonumber \\
c_k &= \frac{\sqrt{g}\gamma_k}{\alpha_k}\, \frac{i\sqrt{q}\tilde g}{V_k g (x_k^{+} + \xi )} 
&
d_k &=  \frac{\sqrt{g}}{\gamma_k}\,\frac{ V_k \tilde g \sqrt{q}(x_k^{+}-x_k^{-})}{ig(\xi x_k^{+}+1)}\,
 \; ,
\end{align}
while $U_k$ and $V_k$ read
\[
\label{u2v2}
U_{k}^{2} = 
q^{-1}\frac{x_k^{+} + \xi}{ x_k^{-} + \xi}
=
q\, \frac{x_k^+}{x_k^-}\,\frac{\xi x_k^{-}+1}{\xi x_k^{+}+1}
\qquad
V_{k}^{2} = 
q^{-1}\frac{\xi x_k^{+}+1}{\xi x_k^{-}+1} 
=
q\,\frac{x_k^+}{x_k^-}\,\frac{x_k^{-}+\xi}{x_k^{+}+\xi} 
\; .
\]

Now the mixed constraints (\ref{FundCond3}) impose a relation between $(x_2^{\pm}, \gamma_2)$ and $(x_4^{\pm}, \gamma_4)$
which is then solved by
\begin{align}
\label{map24}
x_2^{\pm} &= x^{\pm} 
&
\gamma_2 &= \gamma 
\nonumber \\
x_4^{\pm} &= \frac{1}{x^{\pm}}
&
\gamma_4 &=\frac{i\tilde\alpha \gamma}{x^+} \; ,
\end{align}
where the normalization coefficients $\alpha_2$ and $\alpha_4$ are related by (\ref{ParRel}).

A convenient multiplicative evaluation parameter $z$
for our quantum affine algebra turns out to be
\[
\label{eval1}
z=q^{-1}\zeta(x^{+}) = q\zeta(x^{-}) \; .
\]

\paragraph{Cocommutativity.}

The R-matrix of the quantum deformed Hubbard model derived in \cite{Beisert:2008tw} is in fact
invariant under the full quantum affine algebra $\widehat{\halg{Q}}$ defined by the relations (\ref{cmat})--(\ref{coprod}).
More precisely, the cocommutativity relation
\[
\label{coco4}
\rmat \copro(\gen{X}_4) = \coproop(\gen{X}_4) \rmat
\]
is also fulfilled for $\gen{X}_4 \in \set{\gen{K}_4, \gen{E}_4, \gen{F}_4 , \gen{U}_4 , \gen{V}_4}$
in addition to the ones in \eqref{coco}. 

In order to see that it is convenient 
to work with the parametrization in terms of
the variables $x^{\pm}$ and $\gamma$.
Interesting enough 
the relations (\ref{map24}) and (\ref{FundReps2}) state that 
the fundamental representation of $\gen{X}_4$
can be obtained respectively as copies of $\gen{X}_2 \in \set{\gen{K}_2, \gen{E}_2, \gen{F}_2 , \gen{U}_2 , \gen{V}_2}$ 
under the mapping
\[
x^{\pm} \mapsto \frac{1}{x^{\pm}}
\qquad
\gamma \mapsto \frac{i\tilde\alpha \gamma}{x^+} 
\qquad
\alpha \mapsto \alpha\tilde\alpha^2
\qquad
\tilde\alpha \mapsto -\frac{1}{\tilde\alpha}
\;.
\]
Now considering the fundamental R-matrix given in \cite{Beisert:2008tw},%
\footnote{We use a convention for $x^\pm$ which differs slightly from the one used in 
\cite{Beisert:2008tw}, as explained above.} a straightforward computation
reveals that $\rmat$ is invariant under this map up to an overall scalar factor. 
More precisely, $\rmat \mapsto f \rmat$
with some irrelevant scalar factor $f = f (x^{\pm}_1,x^{\pm}_2)$. 
The cocommutativity condition 
for $\gen{X}_2$ in \eqref{coco},
$\rmat \copro(\gen{X}_2) = \coproop(\gen{X}_2) \rmat$,
then directly maps to the one for $\gen{X}_4$ \eqref{coco4}.
This proves the invariance of $\rmat$ under the full quantum affine algebra $\widehat{\halg{Q}}$.

\section{Conventional Quantum Affine Limit}
\label{sec:convlimit}

In this section we aim to investigate the quantum affine algebra $\widehat{\halg{Q}}$ 
and its fundamental representation 
in the limit $g \rightarrow 0$. We shall show that it reduces to the standard
$\halg{U}_q [\widehat{\alg{sl}}(2|2)]$ algebra up to a Reshetikhin twist \cite{Reshetikhin} and a gauge
transformation \cite{Akutsu}.
This limit corresponds to the case ``T(conv)''
in the analysis of the classical algebra \cite{Beisert:2010kk}.

\paragraph{Algebra.}

The affine algebra $\widehat{\halg{Q}}$ differs significantly from the standard $\halg{U}_q [\widehat{\alg{sl}}(2|2)]$ by the fact that 
the anticommutators $\acomm{\gen{E}_2}{\gen{F}_4}$ and $\acomm{\gen{E}_4}{\gen{F}_2}$
do not vanish. Nevertheless, one can readily see from (\ref{psl22}) and (\ref{ParRel}) 
that the above mentioned anticommutators vanish when $g \rightarrow 0$, as well as the central elements deforming the quartic
Serre relations (\ref{qserre}). 
Moreover, in the limit $g\to0$ the relations \eqref{psl22}--\eqref{coprod} almost reproduce the standard products, coproducts, 
antipodes and counits of the quantum affine algebra $\halg{U}_q [\widehat{\alg{sl}}(2|2)]$.

Merely the Hopf algebra structure described in \eqref{coprod} requires a more elaborate analysis. 
The coproducts $\copro(\gen{E}_k)$ and $\copro(\gen{F}_k)$ 
with $k=2,4$ appear twisted by the central elements $\gen{U}_k$
\begin{align}
\label{coprotwist}
\copro(\gen{E}_k)&=\gen{E}_k\otimes 1+\gen{K}_k^{-1} \gen{U}_{2}^{+\delta_{j,2}} \gen{U}_{4}^{+\delta_{k,4}}  \otimes\gen{E}_k
\nln
\copro(\gen{F}_k)&=\gen{F}_k\otimes \gen{K}_k+ \gen{U}_{2}^{-\delta_{j,2}} \gen{U}_{4}^{-\delta_{k,4}} \otimes\gen{F}_k \;.
\end{align}
We recover the standard Hopf algebra structure of the $\halg{U}_q [\widehat{\alg{sl}}(2|2)]$
by the following similarity transformation of the coproduct%
\footnote{We define exponents with coproducts as
$(\gen{U}_2 \otimes 1)^{1\otimes\gen{B}_2}=\exp((\log \gen{U}_2)\otimes\gen{B}_2)$.}
\[
\bar{\copro}(\gen{X}) = 
(\gen{U}_2 \otimes 1)^{-1\otimes\gen{B}_2} 
(\gen{U}_4 \otimes 1)^{-1\otimes\gen{B}_4} 
\copro(\gen{X}) 
(\gen{U}_2 \otimes 1)^{1\otimes\gen{B}_2}
(\gen{U}_4 \otimes 1)^{1\otimes\gen{B}_4}\;,
\]
where $\gen{B}_k$ are two continuous automorphisms of $\halg{U}_q [\widehat{\alg{sl}}(2|2)]$
defined by
\[
\comm{\gen{B}_k}{\gen{E}_j}=+\delta_{j,k}\gen{E}_j
\qquad
\comm{\gen{B}_k}{\gen{K}_j}=0
\qquad
\comm{\gen{B}_k}{\gen{F}_j}=-\delta_{j,k}\gen{F}_j\;.
\]
This clearly removes the central elements $\gen{U}_k$ from the above coproducts
\eqref{coprotwist}.

The above transformation can be viewed as composed from a Reshetikhin twist \cite{Reshetikhin} 
and a change of basis.
The operator 
\[\label{fdef}
\mathcal{F} = 
(1\otimes\gen{U}_2)^{-\gen{B}_2\otimes 1/2} 
(\gen{U}_2\otimes1)^{1\otimes\gen{B}_2/2}
(1\otimes\gen{U}_4)^{-\gen{B}_4\otimes 1/2} 
(\gen{U}_4\otimes1)^{1\otimes\gen{B}_4/2}
\]
satisfies the relations $\mathcal{F}_{12} \mathcal{F}_{21} = 1$ and
$\mathcal{F}_{12} \mathcal{F}_{13} \mathcal{F}_{23} = \mathcal{F}_{23} \mathcal{F}_{13} \mathcal{F}_{12}$.
As demonstrated in \cite{Reshetikhin}, 
$\copro^{(\mathcal{F})}(X)$ and $\rmat^{(\mathcal{F})}$ also form a Hopf algebra with
\[
\label{twist}
\copro^{(\mathcal{F})}(\gen{X}) = \mathcal{F}^{-1} \copro(\gen{X}) \mathcal{F} \qquad \qquad \qquad  
\rmat^{(\mathcal{F})} = \mathcal{F} \rmat \mathcal{F} \; .
\]
The coproduct $\copro^{(\mathcal{F})}$ is already equivalent to the standard coproduct $\bar{\copro}$.
This can be seen upon conjugating the basis 
$\gen{X}'=\gen{U}_2^{-\gen{B}_2/2}\gen{U}_4^{-\gen{B}_4/2}\gen{X}\gen{U}_2^{\gen{B}_2/2}\gen{U}_4^{\gen{B}_4/2}$
which effectively conjugates the coproduct by
\[
(1\otimes\gen{U}_2)^{\gen{B}_2\otimes1/2} 
(\gen{U}_2\otimes 1)^{1\otimes\gen{B}_2/2}
(1\otimes\gen{U}_4)^{\gen{B}_4\otimes1/2} 
(\gen{U}_4\otimes 1)^{1\otimes\gen{B}_4/2}.
\]

\paragraph{Fundamental Representation.}

To understand the limit $g \rightarrow 0$ it is also convenient to consider the fundamental representation of $\widehat{\halg{Q}}$
given in terms of the variables $x^{\pm}$ and $\gamma$. 
Since the variables $x^{+}$ and $x^{-}$ are constrained
by the relation (\ref{eval1}), we first need to introduce an appropriate expansion for them in the proposed limit.
Direct inspection of the relation (\ref{eval1}) leads us to the following expansion,
\[
\label{wce}
x^{\pm} =  \frac{i}{g} \frac{q^{\pm1} z -1}{q - q^{-1}}   + \mathcal{O}(g) 
\qquad \mbox{and} \qquad \gamma = \frac{\bar{\gamma}}{\sqrt{g}} \; ,
\]
where $\bar{\gamma}$ emerges from a rescaling of $\gamma$ required to obtain finite results.

Taking into account the expansion (\ref{wce}), in the limit $g \rightarrow 0$ 
we find that the coefficients $a_k$, $b_k$, $c_k$ and $d_k$ defined
in (\ref{abcd}) assume the following values
\begin{align}
\label{abcd1}
a_2 &= \bar{\gamma}  & b_2 &=  0 
&
a_4 &= 0 & b_4 &=  \alpha \tilde{\alpha} \frac{z}{\bar{\gamma}} 
\nonumber \\
c_2 &= 0 & d_2 &=  \frac{1}{\bar{\gamma}} 
&
c_4 &= -\frac{1}{\alpha \tilde{\alpha} } \frac{\bar{\gamma}}{z} & d_4 &=   0 \;\; .
\end{align}  
Up to some factors these define the canonical representations 
of $\gen{E}_k$, $\gen{F}_k$ in $\halg{U}_q [\widehat{\alg{sl}}(2|2)]$.
In their turn the central element eigenvalues $U_k$ and $V_k$ are then given by
\begin{eqnarray}
\label{uu}
U^2=U_2^{2} = U_4^{-2} = \frac{1-z q }{q-z} \qquad \qquad
V^2=V_2^2 = V_4^{-2} = q \;\; .
\end{eqnarray}
Moreover we find
\[
\label{eval}
\gen{K}_4 \simeq \gen{K}_1^{-1} \gen{K}_2^{-1} \gen{K}_3^{-1}
\qquad
\gen{E}_4 \simeq \alpha \tilde{\alpha} z \bigcomm{\comm{\gen{F}_3}{\gen{F}_2}}{ \gen{F}_1 }
\qquad
\gen{F}_4 \simeq -\alpha^{-1} \tilde{\alpha}^{-1} z^{-1} \left[ \left[\gen{E}_3 ,  \gen{E}_2 \right] , \gen{E}_1 \right] \; ,
\]
which corresponds to the standard evaluation representation 
of the quantum affine algebra $\halg{U}_q [\widehat{\alg{sl}}(2|2)]$ up to a conventional rescaling of the generators
$\gen{E}_4$ and $\gen{F}_4$. 
This observation supports $z$ as the evaluation parameter of the quantum affine algebra $\widehat{\halg{Q}}$.

\paragraph{Fundamental R-Matrix.}

Next we would like to obtain the limit of the fundamental R-matrix.
In order to proceed we need to apply the Reshetikhin twist \eqref{fdef,twist}
to the fundamental R-matrix. 
On the one hand, we have to note that the automorphisms $\gen{B}_2$ and $\gen{B}_4$ 
have no fundamental representation. On the other hand, we are saved by
the fact that they only appear in a combination 
which is represented by the fermion number operator 
\[
\gen{B}=\gen{B}_2-\gen{B}_4\simeq \text{diag}(0,0,1,1)
\]
due to the relation $\gen{U}_2\simeq\gen{U}_4^{-1}$.
Hence the operator $\mathcal{F}$ in \eqref{fdef}
becomes%
\footnote{In the following
$U_i$ denotes the eigenvalue of 
$\gen{U}_2\simeq\gen{U}_4^{-1}$
on site $i$.}
\[
\mathcal{F} \simeq
U_2^{-\gen{B}/2} 
\otimes
U_1^{+\gen{B}/2}.
\]

The matrix elements of $\rmat^{(\mathcal{F})}$ still contain
the factors $\bar{\gamma}_i$ remaining from the normalization
between the bosonic/fermionic states, cf.\ \eqref{abcd1},
as well as some factors of $U_i$.
These can be removed by a spectral parameter dependent gauge transformation \cite{Akutsu}
\[
\bar{\rmat} = ( \mathcal{G}_1 \otimes \mathcal{G}_2 ) \rmat^{(\mathcal{F})} ( \mathcal{G}_1 \otimes \mathcal{G}_2 )^{-1}
\qquad
\mbox{with}
\qquad
\mathcal{G}_i = U_i^{\gen{B}/2} \bar\gamma_i^{-\gen{B}}\;\; .
\]
Altogether the transformation reads
\[
\bar{\rmat} = 
\left[
\bigbrk{\sqrt{U_1/U_2}/\gamma_1}^{\gen{B}}
\otimes
\bigbrk{\sqrt{U_1U_2}/\gamma_2}^{\gen{B}}
\right]
 \rmat 
\left[
\bigbrk{\gamma_1/\sqrt{U_1U_2}}^{\gen{B}}
\otimes
\bigbrk{\gamma_2\sqrt{U_1/U_2}}^{\gen{B}}
\right]\;\;.
\] 
Though we shall not present the explicit form of the R-matrix, 
we find that $\bar{\rmat}$ equals
the R-matrix of the Perk-Schultz model $\halg{U}_q [\widehat{\alg{sl}}(2|2)]$ 
\cite{Perk_Schultz,Schultz,deVega}
up to an overall factor.
Moreover, the matrix $\bar{\rmat}_{ij}$ depends only on the ratio $z_i / z_j$ 
and as expected it satisfies the Yang-Baxter equation in the usual trigonometric form,
\[
\bar{\rmat}_{12}(z_1/z_2)\,
\bar{\rmat}_{13}(z_1/z_3) \,
\bar{\rmat}_{23}(z_2/z_3) 
=
\bar{\rmat}_{23}(z_2/z_3)\,
\bar{\rmat}_{13}(z_1/z_3) \,
\bar{\rmat}_{12}(z_1/z_2) \; .
\]

\section{Yangian Limit}
\label{sec:yangian}

In the previous sections, we have found that the (trigonometric) R-matrix of \cite{Beisert:2008tw}
has a quantum affine symmetry. On the other hand, it is known that the undeformed 
(rational) R-matrix enjoys Yangian symmetry \cite{Beisert:2007ds}. 
Since the quantum deformed fundamental R-matrix (in $x^{\pm}$ parametrization) 
trivially reduces to the undeformed one by taking 
the deformation parameter $q$ to 1, one of the natural questions is 
how the quantum affine symmetry is related to the Yangian symmetry in this limit.  
This is not only an important consistency check of our quantum affine algebra
but also it might serve the possibility to investigate the Yangian structure in the 
AdS/CFT correspondence from the viewpoint of the quantum affine algebra $\widehat{\halg{Q}}$.
This limit corresponds to the case ``R(full)''
in the analysis of the classical algebra \cite{Beisert:2010kk}. 
However, in comparison with the limit of the R-matrix itself, 
the Yangian limit of the quantum affine algebra is not straightforward. 
For instance, if we take the parameter $q$ to 1 naively, 
the quantum affine algebra does not reduce to the Yangian algebra 
but just gives the undeformed universal enveloping algebra. 
Since the Yangian algebra is generated by the level-zero (non-affine) 
and (at least one) level-one generators, we need to find a non-trivial limit 
to obtain the Yangian algebra.

In this section, we show that the AdS/CFT Yangian symmetries \cite{Beisert:2007ds} 
are actually reproduced from our quantum affine algebra $\widehat{\halg{Q}}$. 
The limit is analogous to the Yangian limit of 
the quantum affine $\alg{gl}(n)$ outlined in \appref{sec:glnlimit}.
There is however a subtlety related to an
extra generator of the Yangian $\halg{Y}$,
which was called secret symmetry in \cite{Matsumoto:2007rh}. 
 
\paragraph{Fundamental Representation.}

The difficulty of the Yangian limit in our case is that the affine generators 
$\gen{E}_4$ and $\gen{F}_4$ in \eqref{FundReps2}
do not obey the standard evaluation representation. 
The evaluation representation is helpful to find the algebraic 
identification between the quantum affine algebra and Yangian. 
However we have found that it is possible to take the $q\to1$ limit.
In order to see this, we would like to start with investigating the analytic properties of 
the parameters $a_2,b_2,c_2,d_2$ and $a_4, b_4, c_4, d_4$.
As an important fact, the two sets of parameters are related 
as follows, 
\[
MT_4=
\begin{pmatrix}z^{-1}&0\\0&1\end{pmatrix}
T_2
\begin{pmatrix}w^{-1}&0\\0&wz\end{pmatrix}
\quad\text{with}\quad
M=\begin{pmatrix}0&\alpha \tilde\alpha\\-\alpha^{-1} \tilde\alpha^{-1}&0\end{pmatrix}
\quad
T_k=\begin{pmatrix}a_k&-b_k\\-c_k&d_k\end{pmatrix}
\]
where the evaluation parameter $z$ (cf.\ \eqref{eval1} in $x^\pm$ variables) 
and $w$ are given by
\[
\label{zw}
z=\frac{VU-V^{-1}U^{-1}}{V^{-1}U-VU^{-1}}
\qquad
w=\frac{\tilde g}{g}\frac{q^{1/2}{U}-q^{-1/2}{U}^{-1}}{{V}{U}-{V}^{-1}{U}^{-1}}
=\frac{g}{\tilde g}\frac{{V}^{-1}{U}-{V}{U}^{-1}}{q^{-1/2}{U}-q^{1/2}{U}^{-1}}
\;.
\]

The limit $q\to1$ can be taken in different ways. For the Yangian limit
we assume $U$ to remain finite and arbitrary, 
as expected from \cite{Beisert:2005tm}.
The relation \eqref{FundCond2} between $U$ and $V$ implies
that $V\to 1$. More precisely
as $q=1+h$ for $h\to0$
\[
V=1+hC+\order{h^2}
\qquad
\text{with}
\qquad
C^2=\quarter-g^2(U-U^{-1})^2.
\]
The latter constraint between the central charges $U$ and $C$ 
agrees with \cite{Beisert:2005tm}.
The parameters $x^\pm$ remain finite and they obey the constraint%
\footnote{Even though our $x^\pm$ parametrization is slightly 
different from \cite{Beisert:2008tw}, it has the same $q\to1$ limit.} 
\[
(x^+-x^-)(1-1/x^+x^-)=ig^{-1}
\;.
\]
Using these the central charge eigenvalues take the form familiar form
\[
U^2=\frac{x^+}{x^-}
\qquad
C=\frac{1}{2}\,\frac{1+1/x^+x^-}{1-1/x^+x^-}\;.
\]

It is easy to see that the parameters $z$ and $w$ in \eqref{zw} 
can be expanded as 
\[
z=1-2higu+\mathcal{O}(h^2)
\qquad
w=1+hig(u-v)+\mathcal{O}(h^2)
%
\;.
\]
The rational evaluation parameter $u$ \cite{Beisert:2007ds}
and $v$ are given by  
\begin{align}
\label{uv}
u&=ig^{-1}C\frac{U+U^{-1}}{U-U^{-1}}=\half(x^++x^-)(1+1/x^+x^-)
\nln
v&= ig^{-1}\frac{1}{2}\frac{U+U^{-1}}{U-U^{-1}}=\half(x^++x^-)(1-1/x^+x^-)
\;.
\end{align}

Note that $-\alpha \tilde\alpha (c_4,d_4)\to (a_2,b_2)$
and $\alpha^{-1} \tilde\alpha^{-1}(a_4,b_4)\to (c_2,d_2)$
and hence in the limit $q\to 1$ 
we find
$-\alpha \tilde\alpha \gen{F}_{4}\simeq \gen{E}_{321}$
and 
$\alpha^{-1} \tilde\alpha^{-1} \gen{E}_{4}\simeq \gen{F}_{321}$
with
\[
\gen{E}_{321}=\bigcomm{\comm{\gen{E}_3}{\gen{E}_2}}{\gen{E}_1}
\qquad
\gen{F}_{321}=\bigcomm{\comm{\gen{F}_3}{\gen{F}_2}}{\gen{F}_1}
\;.
\]
That is the limits of $\gen{F}_4$ and $\gen{E}_4$ are not independent 
and the generators should be replaced by the rescaled differences 
$(\alpha\tilde\alpha\gen{F}_{4}+\gen{E}_{321})/(q-1)$
and 
$(\alpha^{-1}\tilde\alpha^{-1}\gen{E}_{4}-\gen{F}_{321})/(q-1)$.
Consequently, what matters in the Yangian limit
is 
\[
\label{paramrel}
\lim_{q\to1}\frac{MT_4-T_2}{ig(q-1)}=
\begin{pmatrix}u&0\\0&-u\end{pmatrix}T_2
+T_2\begin{pmatrix}v&0\\0&-v\end{pmatrix}
=NT_2\;,
\]
where we have introduced the following matrix
\[\label{mnmat}
N
=\begin{pmatrix}2u&-i\alpha(1+{U}^2)\\-i\alpha^{-1}(1+{U}^{-2})&-2u\end{pmatrix}
\;.
\]


\paragraph{Algebra.}
The relation \eqref{paramrel} 
with the matrices \eqref{mnmat}
leads us to the following identification
between the quantum affine algebra $\widehat{\halg{Q}}$ 
and its associated Yangian algebra,
\begin{align}
\label{qtoy}
\lim_{q\to1}\frac{-\alpha\tilde\alpha\gen{F}_{4}-\gen{E}_{321}}{ig(q-1)}
&=2\widehat{\gen{E}}_{321}+i\alpha(1+\gen{U}^2)\gen{F}_2
\nln
\lim_{q\to1}\frac{\alpha^{-1}\tilde\alpha^{-1}\gen{E}_{4}-\gen{F}_{321}}{ig(q-1)}
&=-2\widehat{\gen{F}}_{321}+i\alpha^{-1}(1+\gen{U}^{-2})\gen{E}_2\;,
\end{align}
with the Yangian evaluation representation
\[
\label{yanev}
\widehat{\gen{E}}_{321}\simeq u\gen{E}_{321}\qquad
\widehat{\gen{F}}_{321}\simeq u\gen{F}_{321}\;.
\]
Since the generator $\gen{E}_{321}$ ($\gen{F}_{321}$) is a highest 
(lowest) weight in the adjoint of $\alg{psl}(2|2)$, 
it is sufficient to obtain the other Yangian generators. 
In fact, we have listed all level-one generators in \appref{sec:alllim}.
In comparison with the standard case \eqref{yanlim}, 
the left hand sides of \eqref{qtoy} have the same structure  
but on the right hand sides we need some additional terms. 

The point is that these relations \eqref{qtoy} are actually compatible with the coalgebra structure. 
In other words, the limit of the coproduct on the left hand side of \eqref{qtoy} 
induces the Yangian coproducts on the right hand side, 
\begin{align}
\label{EF321hat}
\lim_{q\to1}\frac{-\alpha\tilde\alpha\copro\gen{F}_{4}-\copro\gen{E}_{321}}{ig(q-1)}
&=\bigl(2\widehat{\gen{E}}_{321}+i\alpha(1+\gen{U}^2)\gen{F}_2-ig^{-1}\gen{k}\gen{E}_{321}\bigr)\otimes1
\nln
&\qquad+\gen{U}\otimes\bigl(2\widehat{\gen{E}}_{321}+i\alpha(1+\gen{U}^2)\gen{F}_2\bigr)
\nln
&\quad-ig^{-1}\big[-\gen{E}_{321}\otimes(\gen{H}_3+\gen{H}_2+\gen{H}_1)
+(\gen{H}_3+\gen{H}_2+\gen{H}_1)\gen{U}\otimes\gen{E}_{321}
\nln
&\qquad+\gen{E}_{32}\otimes\gen{E}_1-\gen{E}_1\gen{U}\otimes\gen{E}_{32}
-\gen{E}_3\gen{U}\otimes\gen{E}_{21}+\gen{E}_{21}\otimes\gen{E}_3\big]
\nln
&=\copro\bigl(2\widehat{\gen{E}}_{321}+i\alpha(1+\gen{U}^2)\gen{F}_2\bigr)
\nln
\lim_{q\to1}\frac{\alpha^{-1}\tilde\alpha^{-1}\copro\gen{E}_4-\copro\gen{F}_{321}}{ig(q-1)}
&=\bigl(-2\widehat{\gen{F}}_{321}+i\alpha^{-1}(1+\gen{U}^{-2})\gen{E}_2\bigr)\otimes1
\nln
&\qquad+\gen{U}^{-1}\otimes\bigl(-2\widehat{\gen{F}}_{321}
+i\alpha^{-1}(1+\gen{U}^{-2})\gen{E}_2+ig^{-1}\gen{k}\gen{F}_{321}\bigr)
\nln
&\quad+ig^{-1}\bigl[\gen{F}_{321}\otimes(\gen{H}_3+\gen{H}_2+\gen{H}_1)
-(\gen{H}_3+\gen{H}_2+\gen{H}_1)\gen{U}^{-1}\otimes\gen{F}_{321}
\nln
&\qquad+\gen{F}_{32}\otimes\gen{F}_1-\gen{F}_1\gen{U}^{-1}\otimes\gen{F}_{32}
-\gen{F}_3\gen{U}^{-1}\otimes\gen{F}_{21}+\gen{F}_{21}\otimes\gen{F}_3\bigr]
\nln
&=\copro\bigl(-2\widehat{\gen{F}}_{321}+i\alpha^{-1}(1+\gen{U}^{-2})\gen{E}_2\bigr)
\;.
\end{align}
Here $\gen{k}$ is the affine central element given by 
$\gen{k}=\gen{H}_1+\gen{H}_2+\gen{H}_3+\gen{H}_4$
whose eigenvalue $k$ vanishes on evaluation representations.
These coproducts coincide with the results in \cite{Beisert:2007ds}
where $\gen{k}$ was projected out,
cf.\ \appref{sec:alllim} for a translation of the generator notation.
Note that the additional terms 
$i\alpha(1+\gen{U}^2)\gen{F}_2$ and $i\alpha^{-1}(1+\gen{U}^{-2})\gen{E}_2$ 
on the right hand side of \eqref{EF321hat} are required to 
cancel certain contributions from the higher central charges in
the original Yangian symmetries $\copro\widehat{\gen{E}}_{321}$ 
and $\copro\widehat{\gen{F}}_{321}$ \cite{Beisert:2007ds}.

The deeper meaning of the additional terms in \eqref{qtoy}
is not clear to us. It is nevertheless interesting to interpret them
as a contribution of an extra Yangian generator $\widehat{\gen{B}}$
called secret symmetry \cite{Matsumoto:2007rh}.
The fundamental representation of this generator is given by
\[
\label{secret}
\widehat{\gen{B}}\simeq\frac{v}{2}\,\mbox{diag}(1,1,-1,-1)
\]
with the parameter $v$ in \eqref{uv}. 
The relevant two of its commutators read \cite{Beisert:2007ty}
\[
\comm{\widehat{\gen{B}}}{\gen{E}_{321}}
=-\widehat{\gen{E}}_{321}-i\alpha(1+\gen{U}^2)\gen{F}_2
\qquad
\comm{\widehat{\gen{B}}}{\gen{F}_{321}}
=\widehat{\gen{F}}_{321}-i\alpha^{-1}(1+\gen{U}^{-2})\gen{E}_2\;.
\]
These are indeed compatible with their coproducts, 
therefore the equivalent replacement in \eqref{EF321hat} is valid as well.
Using this secret symmetry, we can rewrite the Yangian limit \eqref{qtoy} as 
\[
\lim_{q\to1}\frac{-\alpha\tilde\alpha\gen{F}_4-\gen{E}_{321}}{ig(q-1)}
=\widehat{\gen{E}}_{321}-\comm{\widehat{\gen{B}}}{\gen{E}_{321}}
\qquad
\lim_{q\to1}\frac{\alpha^{-1}\tilde\alpha^{-1}\gen{E}_4-\gen{F}_{321}}{ig(q-1)}
=-\widehat{\gen{F}}_{321}-\comm{\widehat{\gen{B}}}{\gen{F}_{321}}\;.
\]

\section{Conclusions}
\label{sec:concl}

In this work we have derived a novel quantum affine algebra $\widehat{\mathcal{Q}}$ 
based on a central extension of the $\alg{sl}(2|2)$ Lie superalgebra. 
As a matter of fact this algebra emerges naturally from compatibility
requirements with the R-matrix of the deformed Hubbard chain 
\cite{Beisert:2008tw} also known as the Alcaraz-Bariev model \cite{alcbar}.
In this sense the formulation of this algebra sheds some new light into a more 
complete understanding of the integrable structure underlying the Hubbard model 
and its deformed counterpart.

The construction of the quantum affine algebra $\widehat{\mathcal{Q}}$ was immensely guided by the Dynkin diagram
of the $\widehat{\alg{sl}}(2|2)$ algebra. More precisely, the similarity between the fermionic nodes 2 and 4 of the Dynkin diagram
given in \figref{fig:OXOX} suggests for instance that the generators associated to the node 4 should act as 
copies of ones associated to the node 2. This observation has played a fundamental role not only for the 
establishment of the commutation relations (\ref{psl22}), 
but also for the construction of the fundamental representation. 

The quantum affine algebra $\widehat{\mathcal{Q}}$ possesses fundamentally a deformation parameter $q$ originated from
the deformation of the universal enveloping algebra of $\alg{sl}(2|2)$, as well as a coupling parameter $g$ introduced
by the central extensions. 
Here we have also shown that the algebra $\widehat{\mathcal{Q}}$ reduces to the standard quantum affine algebra 
$\mathcal{U}_{q}[\widehat{\alg{sl}}(2|2)]$ in the limit $g \rightarrow 0$, which unveils a relation between
the Alcaraz-Bariev model and the Perk-Schultz model $\mathcal{U}_{q}[\widehat{\alg{sl}}(2|2)]$ in this particular limit.
We have furthermore investigated the limit $q \rightarrow 1$ where we have found that the affine algebra $\widehat{\mathcal{Q}}$
reproduces the Yangian $\mathcal{Y}$ of a centrally extended $\alg{sl}(2|2)$ algebra. This Yangian $\mathcal{Y}$ corresponds
to the same algebra underlying Shastry's R-matrix which also plays an important role for integrability in the context of the AdS/CFT
correspondence. In this way, as quantum affine algebras offer a more uniform description
in comparison to Yangians, this limit procedure might help us address 
integrability in the AdS/CFT correspondence.
 
In the analysis of the classical algebra performed in \cite{Beisert:2010kk}, 
the conventional quantum affine and Yangian limits reduces respectively
to the cases ``T(conv)'' and ``R(full)''. 
However, in the classical limit a 
whole cascade of algebras has been presented in \cite{Beisert:2010kk} 
which make us wonder if all the cases indeed possess a quantum counterpart. 

Furthermore it would be worthwhile
to investigate higher representations of the algebra,
cf.\ \cite{Zhang:2004qx},
which are likely to be direct analogs of the
undeformed case studied in 
\cite{Beisert:2006qh,Chen:2006gp,Arutyunov:2009mi,Arutyunov:2009pw}.
Finally, the formulation of Drinfel'd's second realization
for this algebra would constitute a valuable step towards 
the universal R-matrix, cf.\ \cite{Spill:2008tp}.

\paragraph{Acknowledgements.}

The authors would like to warmly thank
M.\ de\ Leeuw, 
L.\ Gow,
H.\ Kanno,
T.\ McLoughlin, 
S.\ Moriyama
and 
H.\ Shimada
for interesting discussions.
%
%
T.M.\ thanks Max-Planck-Institut f\"ur Gravitationsphysik,
Albert-Einstein-Institut, in Potsdam for the hospitality 
where main part of this work was performed. 
The work of T.M.\ is supported in part by 
Japan Society for the Promotion of Science(JSPS)
Grant-in-Aid for JSPS Fellows  No.21$\cdot$4809 
and ``Excellent Young Researcher Overseas Visit Program.'' 
The work of N.B.\ is supported in part by 
the German-Israeli Foundation (GIF).

\appendix

\section{Yangian Limit of Quantum Affine \texorpdfstring{$\alg{gl}(n)$}{gl(n)}}
\label{sec:glnlimit}
As we have mentioned in the beginning of \secref{sec:yangian}, 
the Yangian limit is not so much trivial. 
Therefore it is convenient to review the generic example of 
$\alg{gl}(n)$ case ($n\geq 3$) 
\cite{2002math.....12370T, 1998math......3008T, 2001PAN....64.2179T}.
That is the limit from $\halg{U}_q[\widehat{\alg{gl}}(n)]$ to $\halg{Y}[\alg{gl}(n)]$, 
which enable us to make the logic clear. 

The generators of the Lie algebra $\alg{gl}(n)$ are given by $\gen{J}^i{}_j$ 
with $i,j=1,\cdots,n$ and they satisfy the standard commutation relations,
\[
\comm{\gen{J}^i{}_j}{\gen{J}^k{}_l}=\delta^k_j\gen{J}^i{}_l-\delta^i_l\gen{J}^j{}_k
\;.
\]

In order to describe its quantum deformation $\halg{U}_q[\alg{gl}(n)]$, 
it is convenient to introduce the corresponding Chevalley-Serre simple roots 
$\gen{E}_i,\gen{F}_i,\gen{H}_i$ with $i,j=1,\cdots,n-1$, which are related as 
\[
\gen{E}_i=\gen{J}^i{}_{i+1} 
\qquad 
\gen{F}_i=\gen{J}^{i+1}{}_i 
\qquad 
\gen{H}_i=\gen{J}^i{}_i-\gen{J}^{i+1}{}_{i+1}
\;.
\]
Their commutation relations are given by 
\[\label{gl}
\comm{\gen{H}_i}{\gen{E}_j}=+A_{ij}\gen{E}_j
\qquad
\comm{\gen{H}_i}{\gen{F}_j}=-A_{ij}\gen{F}_j
\qquad
\comm{\gen{E}_i}{\gen{F}_j}=\delta_{ij}
\frac{q^{\gen{H}_i}-q^{-\gen{H}_i}}{q-q^{-1}}
\]
with the Cartan matrix $A$ defined by 
\[\label{glcartan}
A_{ij}=
\begin{cases}
+2&\text{for } i=j
\\
-1&\text{for } |i-j|=1
\\
0&\text{for } |i-j|\geq 2
\;.
\end{cases}
\]
Furthermore, the following Serre relations hold for $|i-j|=1$
\begin{align}\label{glserre}
\bigcomm{\gen{E}_i}{\comm{\gen{E}_i}{\gen{E}_j}}
&=(q-2+q^{-1})\gen{E}_i\gen{E}_j\gen{E}_i
\nln
\bigcomm{\gen{F}_i}{\comm{\gen{F}_i}{\gen{F}_j}}
&=(q-2+q^{-1})\gen{F}_i\gen{F}_j\gen{F}_i
\end{align}
and for $|i-j|\geq2$
\[\label{glserre2}
\comm{\gen{E}_i}{\gen{E}_j}
=\comm{\gen{F}_i}{\gen{F}_j}=0 
\;.
\]

The affine extension $\halg{U}_q[\widehat{\alg{gl}}(n)]$ to 
$\halg{U}_q[\alg{gl}(n)]$
is obtained by adding the affine generators $\gen{E}_n,\gen{F}_n,\gen{H}_n$
and extending the Cartan matrix to $n\times n$.
The relations are almost same as the above but the indices in \eqref{gl},
\eqref{glcartan}, \eqref{glserre} and \eqref{glserre2} are considered modulo $n$.
It is noted that the summation of the Cartan generators 
$\gen{H}_1+\cdots+\gen{H}_n=k$ turns out to be the affine central element. 

The quantum affine algebra has a Hopf algebra structure.
For the the Chevalley-Serre generators, the coproducts, antipodes and 
counits are given by, 
\begin{align}
\copro(\gen{E}_i)&=\gen{E}_i\otimes1+q^{-\gen{H}_i}\otimes\gen{E}_i
&
\antipode(\gen{E}_i)&=-q^{\gen{H}_i}\gen{E}_i
&
\counit(\gen{E}_i)&=0
\nln
\copro(\gen{F}_i)&=\gen{F}_i\otimes q^{\gen{H}_i}+1\otimes\gen{F}_i
&
\antipode(\gen{F}_i)&=-\gen{F}_iq^{-\gen{H}_i}
&
\counit(\gen{F}_i)&=0
\nln
\copro(\gen{H}_i)&=\gen{H}_i\otimes1+1\otimes\gen{H}_i
&
\antipode(\gen{H}_i)&=-\gen{H}_i
&
\counit(\gen{H}_i)&=0
\;.
\end{align}

One of the important representations of the algebra is the evaluation representation,
in which the affine generators are expressed as
\[
\label{glnev}
\gen{E}_n\simeq z^{-1}q^{\gen{J}^1{}_1+\gen{J}^n{}_n}\gen{F}_{n-1\cdots1}\qquad
\gen{F}_n\simeq z q^{-\gen{J}^1{}_1-\gen{J}^n{}_n}\gen{E}_{n-1\cdots1}\qquad
\gen{H}_n\simeq -\gen{H}_{n-1}-\cdots-\gen{H}_1
\]
with the evaluation parameter $z$. Here we have used the following abbreviations
\[
\label{short}
\gen{E}_{n-1\cdots1}=[[\comm{\gen{E}_{n-1}}{\gen{E}_{n-2}}_q,\cdots]_q,\gen{E}_1]_q
\quad\;
\gen{F}_{n-1\cdots1}=[[\comm{\gen{F}_{n-1}}{\gen{F}_{n-2}}_{q^{-1}},\cdots]_{q^{-1}},\gen{F}_1]_{q^{-1}}
\;,
\]
where the $q$-deformed commutators are defined by 
\[
\comm{\gen{A}}{\gen{B}}_{q^{\pm1}}=\gen{A}\gen{B}-q^{\pm1}\gen{B}\gen{A}
\;.
\]
Noting that the affine central element $\gen{k}$ vanishes in this representation. 

The Yangian limit is taken by the following identification,
\[
\label{yanlim}
\lim_{q\to1}\frac{\gen{F}_n-q^{-\gen{J}^1{}_1-\gen{J}^n{}_n}\gen{E}_{n-1\cdots1}}{q-1}
=\widehat{\gen{E}}_{n-1\cdots1}
\qquad
\lim_{q\to1}\frac{q^{\gen{J}^1{}_1+\gen{J}^n{}_n}\gen{F}_{n-1\cdots1}-\gen{E}_n}{q-1}
=\widehat{\gen{F}}_{n-1\cdots1}
\;.
\]
The left hand side of the above relations are the $q\to1$ limit of the quantum affine 
generators and the right hand sides are the level-one Yangian generators. 
This identification \eqref{yanlim} has two good properties. 
The first one is the consistency with the Yangian evaluation representation,
\[
\widehat{\gen{E}}_{n-1\cdots1}\simeq u\gen{E}_{n-1\cdots1}
\qquad
\widehat{\gen{F}}_{n-1\cdots1}\simeq u\gen{F}_{n-1\cdots1}
\]
where the Yangian evaluation parameter $u$ is related with the quantum one 
in \eqref{glnev} as $z=q^u$
and the bookkeeping notations \eqref{short} are replaced by $q=1$. 
The second one is the compatibility with the coproducts. In other words, the following 
Yangian coproducts are automatically derived from the quantum affine algebra from 
the relations \eqref{yanlim} up to the affine central element $\gen{k}$,
\begin{align}
\copro\widehat{\gen{E}}_{n-1\cdots1}
&=\bigl(\widehat{\gen{E}}_{n-1\cdots1}+\gen{k}\gen{E}_{n-1\cdots1}\bigr)\otimes1
+1\otimes\widehat{\gen{E}}_{n-1\cdots1}
\nln
&\quad
+2\bigl[
\gen{E}_{n-1\cdots1}\otimes\gen{J}^n{}_n+\gen{J}^1{}_1\otimes\gen{E}_{n-1\cdots1}
+\sum^{n-2}_{k=1}\gen{E}_{n-1\cdots k+1}\otimes\gen{E}_{k\cdots 1}
\bigr]
\nln
\copro\widehat{\gen{F}}_{n-1\cdots1}
&=\widehat{\gen{F}}_{n-1\cdots1}\otimes1
+1\otimes\bigl(\widehat{\gen{F}}_{n-1\cdots1}+\gen{k}\gen{F}_{n-1\cdots1}\bigr)
\nln
&\quad
+2\bigl[
\gen{F}_{n-1\cdots1}\otimes\gen{J}^1{}_1+\gen{J}^n{}_n\otimes \gen{F}_{n-1\cdots1}
-\sum^{n-2}_{k=1}\gen{F}_{1\cdots k}\otimes\gen{F}_{k+1\cdots n-1}
\bigr]
\;.
\end{align}
In fact, the defining relations of the Yangian algebra $\halg{Y}[\alg{gl}(n)]$ stem from those of 
the quantum affine algebra $\halg{U}_q[\widehat{\alg{gl}}(n)]$ via the identification \eqref{yanlim}.

\section{Yangian Limits for All Generators}
\label{sec:alllim}

In this appendix, we would like to list the Yangian limits for all the generators 
in the quantum affine algebra $\halg{Q}$ for the completeness. 
In order to do that, it is convenient to introduces some notations
$\gen{Q}^{a\alpha}=\epsilon^{ab}\gen{Q}^\alpha{}_b$ and 
$\gen{S}^{a\alpha}=\epsilon^{\alpha\beta}\gen{S}^a{}_\beta$ ($a,\alpha=1,2$) 
for the fermionic generators \cite{Beisert:2005tm, Beisert:2006qh}. 
These generators are defined by Chevalley-Serre basis as
\begin{align}
\label{eq:QSdef}
\gen{Q}^{11}&=\gen{E}_{32}&
\gen{Q}^{12}&=\gen{E}_{2}&
\gen{Q}^{21}&=-\gen{E}_{321}&
\gen{Q}^{22}&=-\gen{E}_{21}
\nln
\gen{S}^{11}&=-\gen{F}_{21}&
\gen{S}^{12}&=-\gen{F}_{321}&
\gen{S}^{21}&=\gen{F}_{2}&
\gen{S}^{22}&=\gen{F}_{32}
\;.
\end{align}
We also denote another set of fermionic generators which include the affine 
generators $\gen{E}_4, \gen{F}_4$ as 
$\overline{\gen{Q}}{}^\alpha{}_a, \overline{\gen{S}}{}^a{}_\alpha$.
They are given by replacing $\gen{E}_2, \gen{F}_2$ to $\gen{E}_4, \gen{F}_4$ 
in \eqref{eq:QSdef} respectively,  
\begin{align}
\overline{\gen{Q}}{}^{11}&=\gen{E}_{34}&
\overline{\gen{Q}}{}^{12}&=\gen{E}_{4}&
\overline{\gen{Q}}{}^{21}&=-\gen{E}_{341}&
\overline{\gen{Q}}{}^{22}&=-\gen{E}_{41}
\nln
\overline{\gen{S}}{}^{11}&=-\gen{F}_{41}&
\overline{\gen{S}}{}^{12}&=-\gen{F}_{341}&
\overline{\gen{S}}{}^{21}&=\gen{F}_{4}&
\overline{\gen{S}}{}^{22}&=\gen{F}_{34}
\;.
\end{align}
This notation allows us to express the Yangian limits in synthesized forms. 
The Yangian limits for the fermionic generators are now given by
\begin{align}
\label{QSlimit}
\lim_{q\to1}\frac{\alpha\tilde\alpha \overline{\gen{S}}{}^{a\alpha}-\gen{Q}^{a\alpha}}{ig(q-1)}
&=2\widehat{\gen{Q}}^{a\alpha}-i\alpha(1+\gen{U}^2)\gen{S}^{a\alpha}
=\widehat{\gen{Q}}^{a\alpha}-\comm{\widehat{\gen{B}}}{\gen{Q}^{a\alpha}}
\nln
\lim_{q\to1}\frac{\alpha^{-1}\tilde\alpha^{-1} \overline{\gen{Q}}{}^{a\alpha}
+
\gen{S}^{a\alpha}}{ig(q-1)}
&=2\widehat{\gen{S}}^{a\alpha}+i\alpha^{-1}(1+\gen{U}^{-2})\gen{Q}^{a\alpha}
=\widehat{\gen{S}}^{a\alpha}
+\comm{\widehat{\gen{B}}}{\gen{S}^{a\alpha}}
\;.
\end{align}
The other Yangian limit for the bosonic generators, which are defined by
\begin{align}
&\gen{R}^{11}=-\gen{F}_{1}&
&\gen{R}^{12}=\gen{R}^{21}=-\half\gen{H}_{1}&
&\gen{R}^{22}=\gen{E}_{1}
\nln
&\gen{L}^{11}=-\gen{E}_{3}&
&\gen{L}^{12}=\gen{L}^{21}=-\half\gen{H}_{3}&
&\gen{L}^{22}=\gen{F}_{3}
\nln
&\gen{C}=-\tfrac{1}{2}\gen{H}_1-\gen{H}_2-\tfrac{1}{2}\gen{H}_3&
&\gen{P}=\bigacomm{\comm{\gen{E}_1}{\gen{E}_2}}{\comm{\gen{E}_3}{\gen{E}_2}}&
&\gen{K}=\bigacomm{\comm{\gen{F}_1}{\gen{F}_2}}{\comm{\gen{F}_3}{\gen{F}_2}}
\;,
\end{align}
are inductively obtained by computing suitable commutation relations
from \eqref{QSlimit} as 
\begin{align}
\label{RLClim}
&\lim_{q\to1}\frac{\alpha\tilde\alpha\acomm{\overline{\gen{S}}{}^{a\alpha}}
{\gen{Q}^{b\beta}}-\epsilon^{ab}\epsilon^{\alpha\beta}\gen{P}}{2ig(q-1)}
=\epsilon^{ab}\epsilon^{\alpha\beta}\widehat{\gen{P}}
+\tfrac{i}{2}\alpha(1+\gen{U}^2)
(\epsilon^{\alpha\beta}\gen{R}^{ab}-\epsilon^{ab}\gen{L}^{\alpha\beta}+\epsilon^{ab}\epsilon^{\alpha\beta}\gen{C})
\nln
&\lim_{q\to1}\frac{\acomm{\overline{\gen{Q}}{}^{a\alpha}}{\gen{S}^{b\beta}}
+\alpha\tilde\alpha \epsilon^{ab}\epsilon^{\alpha\beta}\gen{K}}{2\alpha\tilde\alpha ig(q-1)}
=\epsilon^{ab}\epsilon^{\alpha\beta}\widehat{\gen{K}}
+\tfrac{i}{2}\alpha^{-1}(1+\gen{U}^{-2})
(\epsilon^{\alpha\beta}\gen{R}^{ab}-\epsilon^{ab}\gen{L}^{\alpha\beta}-\epsilon^{ab}\epsilon^{\alpha\beta}\gen{C})
\nln
&\lim_{q\to1}\frac{\alpha^{-1}\tilde\alpha^{-1}\acomm{\overline{\gen{Q}}{}^{a\alpha}}{\gen{Q}^{b\beta}}
+\alpha\tilde\alpha\acomm{\gen{S}^{a\alpha}}{\overline{\gen{S}}{}^{b\beta}}}{4ig(q-1)}
\nln
&
\qquad \qquad \qquad \qquad \qquad \qquad 
=
-\epsilon^{\alpha\beta}\widehat{\gen{R}}{}^{ab}
+\epsilon^{ab}\widehat{\gen{L}}{}^{\alpha\beta}
-\epsilon^{ab}\epsilon^{\alpha\beta}\widehat{\gen{C}}
-\tfrac{i}{2}g\epsilon^{ab}\epsilon^{\alpha\beta}(U^2-U^{-2})
\;.
\end{align}
The above limits \eqref{QSlimit} and \eqref{RLClim} give the same coproducts 
presented by \cite{Beisert:2007ds, Matsumoto:2007rh} and the symmetries of the undeformed 
R-matrix \cite{Beisert:2005tm}.

\begin{bibtex}
@article{1998math......3008T,
  author = {Tolstoy, V. N.},
  title = {Drinfeldians},
  archiveprefix = {arXiv},
  eprint = {math.QA/9803008},
  slaccitation = {
  warnincomplete = {WARNING},
}

@article{2001PAN....64.2179T,
  author = {Tolstoy, V. N.},
  title = {Super-Drinfeldians and Super-Yangians of Lie Superalgebras of Type A(n$/$m)},
  year = {2001},
  journal = {Physics of Atomic Nuclei},
  volume = {64},
  pages = {2179-2184},
  doi = {10.1134/1.1432922},
}

@article{2002math.....12370T,
  author = {Tolstoy, V. N.},
  title = {From quantum affine Kac-Moody algebras to Drinfeldians and Yangians},
  archiveprefix = {arXiv},
  eprint = {math.QA/0212370},
  slaccitation = {
  journal = {Contemporary Mathematics},
  volume = {343},
  pages = {349},
  year = {2004},
  warnincomplete = {WARNING},
}

@article{Ahn:2010ka,
  author = {Ahn, Changrim and Nepomechie, Rafael I.},
  title = {Review of AdS/CFT Integrability, Chapter III.2: Exact world-sheet S-matrix},
      journal        = "Lett.Math.Phys.",
      volume         = "99",
      pages          = "209-229",
      doi            = "10.1007/s11005-011-0478-9",
      year           = "2012",
  archiveprefix = {arXiv},
  eprint = {1012.3991},
  primaryclass = {hep-th},
  slaccitation = {
  warnincomplete = {WARNING},
}

@article{Akutsu,
  author = {Wadati, M. and Deguchi, T. and Akutsu, Y.},
  title = {Exactly solvable models and knot theory},
  year = {1989},
  journal = {Phys. Rept.},
  volume = {180},
  pages = {247},
  doi = {10.1016/0370-1573(89)90123-3},
}

@article{alcbar,
  author = {Alcaraz, F. C. and Bariev, R. Z.},
  title = {Interpolation between Hubbard and supersymmetric t-J models: two-parameter integrable models of correlated electrons},
  year = {1999},
  journal = {J. Phys.},
  volume = {A32},
  pages = {L483},
  doi = {10.1088/0305-4470/32/46/101},
  archiveprefix = {arXiv},
  eprint = {cond-mat/9908265},
  slaccitation = {
}

@article{Artin,
  author = {Artin, E.},
  title = {Theory of braids},
  year = {1947},
  journal = {Ann. Math.},
  volume = {48},
  pages = {101},
  warnincomplete = {WARNING},
}

@article{Arutyunov:2009pw,
  author = {Arutyunov, Gleb and de Leeuw, Marius and Torrielli, Alessandro},
  title = {On Yangian and Long Representations of the Centrally Extended su(2$/$2) Superalgebra},
  year = {2010},
  journal = {JHEP},
  volume = {1006},
  pages = {033},
  doi = {10.1007/JHEP06(2010)033},
  archiveprefix = {arXiv},
  eprint = {0912.0209},
  primaryclass = {hep-th},
  slaccitation = {
}

@article{Bazh2,
  author = {Bazhanov, V. V. and Shadrikov, A. G.},
  title = {Trigonometric solutions of triangle equations - simple Lie superalgebras},
  year = {1987},
  journal = {Theor. Math. Phys.},
  volume = {73},
  pages = {1302},
  doi = {10.1007/BF01041913},
  warnincomplete = {WARNING},
}

@article{Bazhanov:1984gu,
  author = {Bazhanov, V. V.},
  title = {Trigonometric Solution Of Triangle Equations And Classical Lie Algebras},
  year = {1985},
  journal = {Phys. Lett.},
  volume = {B159},
  pages = {321-324},
  doi = {10.1016/0370-2693(85)90259-X},
  slaccitation = {
}

@article{Beisert:2005tm,
  author = {Beisert, Niklas},
  title = {The su(2$/$2) dynamic S-matrix},
  year = {2008},
  journal = {Adv. Theor. Math. Phys.},
  volume = {12},
  pages = {945},
  archiveprefix = {arXiv},
  eprint = {hep-th/0511082},
  slaccitation = {
}

@article{Beisert:2006qh,
  author = {Beisert, Niklas},
  title = {The Analytic Bethe Ansatz for a Chain with Centrally Extended su(2$/$2) Symmetry},
  year = {2007},
  journal = {J. Stat. Mech.},
  volume = {07},
  pages = {P01017},
  doi = {10.1088/1742-5468/2007/01/P01017},
  archiveprefix = {arXiv},
  eprint = {nlin.SI/0610017},
  slaccitation = {
}

@article{Beisert:2007ds,
  author = {Beisert, Niklas},
  title = {The S-Matrix of AdS/CFT and Yangian Symmetry},
  year = {2007},
  journal = {PoS},
  volume = {Solvay},
  pages = {002},
  archiveprefix = {arXiv},
  eprint = {0704.0400},
  primaryclass = {nlin.SI},
  slaccitation = {
  warnincomplete = {WARNING},
}

@article{Beisert:2007ty,
  author = {Beisert, Niklas and Spill, Fabian},
  title = {The Classical r-matrix of AdS/CFT and its Lie Bialgebra Structure},
  year = {2009},
  journal = {Commun. Math. Phys.},
  volume = {285},
  pages = {537-565},
  doi = {10.1007/s00220-008-0578-2},
  archiveprefix = {arXiv},
  eprint = {0708.1762},
  primaryclass = {hep-th},
  slaccitation = {
}

@article{Beisert:2008tw,
  author = {Beisert, Niklas and Koroteev, Peter},
  title = {Quantum Deformations of the One-Dimensional Hubbard Model},
  year = {2008},
  journal = {J. Phys.},
  volume = {A41},
  pages = {255204},
  doi = {10.1088/1751-8113/41/25/255204},
  archiveprefix = {arXiv},
  eprint = {0802.0777},
  primaryclass = {hep-th},
  slaccitation = {
}

@article{Beisert:2010jr,
  author = {Beisert, Niklas and others},
  title = {Review of AdS/CFT Integrability: An Overview},
      journal        = "Lett.Math.Phys.",
      volume         = "99",
      pages          = "3-32",
      doi            = "10.1007/s11005-011-0529-2",
      year           = "2012",
  archiveprefix = {arXiv},
  eprint = {1012.3982},
  primaryclass = {hep-th},
  slaccitation = {
  warnincomplete = {WARNING},
}

@article{Beisert:2010kk,
  author = {Beisert, Niklas},
  title = {The Classical Trigonometric r-Matrix for the Quantum-Deformed Hubbard Chain},
      journal        = "J.Phys.",
      volume         = "A44",
      pages          = "265202",
      doi            = "10.1088/1751-8113/44/26/265202",
      year           = "2011",
  archiveprefix = {arXiv},
  eprint = {1002.1097},
  primaryclass = {math-ph},
  slaccitation = {
  warnincomplete = {WARNING},
}

@article{Bernard:1990ys,
  author = {Bernard, Denis and Leclair, Andre},
  title = {Quantum group symmetries and nonlocal currents in 2-D QFT},
  year = {1991},
  journal = {Commun. Math. Phys.},
  volume = {142},
  pages = {99-138},
  doi = {10.1007/BF02099173},
  slaccitation = {
}

@article{chari_pressley,
  author = {Chari, V. and Pressley, A.},
  title = {Small representations of quantum affine algebras},
  year = {1994},
  journal = {Lett. Math. Phys.},
  volume = {30},
  pages = {131},
  doi = {10.1007/BF00939701},
}

@article{chari_pressley1,
  author = {Chari, V. and Pressley, A.},
  title = {Quantum affine algebras and their representations},
  year = {1995},
  journal = {Canadian Mathematical Society Conference Proceedings},
  volume = {16},
  pages = {59},
  archiveprefix = {arXiv},
  eprint = {hep-th/9411145},
  primaryclass = {hep-th},
  warnincomplete = {WARNING},
}

@article{chari_pressley1a,
  author = {Chari, V. and Pressley, A.},
  title = {Quantum affine algebras},
  year = {1991},
  journal = {Commun. Math. Phys.},
  volume = {142},
  pages = {261},
  doi = {10.1007/BF02102063},
}

@article{chari_pressley2,
  author = {Chari, V. and Pressley, A.},
  title = {New unitary representations of loop groups},
  year = {1986},
  journal = {Math. Ann.},
  volume = {275},
  pages = {87},
  doi = {10.1007/BF01458586},
}

@article{deVega,
  author = {de Vega, H. J. and Lopes, E.},
  title = {Exact solution of the Perk-Schultz model},
  year = {1991},
  journal = {Phys. Rev. Lett.},
  volume = {67},
  pages = {489},
  doi = {10.1103/PhysRevLett.67.489},
}

@article{drinfeld,
  author = {Drinfel'd, V. G.},
  title = {Hopf algebras and the quantum Yang-Baxter equation},
  year = {1985},
  journal = {Sov. Math. Dokl.},
  volume = {32},
  pages = {254-258},
  slaccitation = {
  warnincomplete = {WARNING},
}

@book{Essler:2005aa,
  author = {Essler, F. H. L. and Frahm, H. and G\"ohmann, F. and Kl\"umper, A. and Korepin, V. E.},
  title = {The one-dimensional Hubbard model},
  year = {2005},
  address = {Cambridge, UK},
  publisher = {Cambridge University Press},
}

@article{Galleas:2006kd,
  author = {Galleas, W. and Martins, M. J.},
  title = {New R-matrices from representations of braid-monoid algebras based on superalgebras},
  year = {2006},
  journal = {Nucl. Phys.},
  volume = {B732},
  pages = {444-462},
  doi = {10.1016/j.nuclphysb.2005.10.025},
  archiveprefix = {arXiv},
  eprint = {nlin/0509014},
  slaccitation = {
}

@article{Ge,
  author = {Ge, M. L. and Wu, Y. S. and Xue, K.},
  title = {Explicit Trigonometric Yang-Baxterization},
  year = {1991},
  journal = {Int. J. Mod. Phys.},
  volume = {A6},
  pages = {3735},
  doi = {10.1142/S0217751X91001817},
}

@article{Gomez:2006va,
  author = {G\'omez, C\'esar and Hern\'andez, Rafael},
  title = {The magnon kinematics of the AdS/CFT correspondence},
  year = {2006},
  journal = {JHEP},
  volume = {0611},
  pages = {021},
  doi = {10.1088/1126-6708/2006/11/021},
  archiveprefix = {arXiv},
  eprint = {hep-th/0608029},
  slaccitation = {
}

@article{heilmann,
  author = {Heilmann, O. J. and Lieb, E. H.},
  title = {Violation of the noncrossing rule: The Hubbard hamiltonian for benzene},
  year = {1971},
  journal = {Annals of the New York Academy of Sciences},
  volume = {172},
  pages = {584},
  warnincomplete = {WARNING},
}

@article{Iohara:2001aa,
  author = {Iohara, Kenji and Koga, Yoshiyuki},
  title = {Central extensions of Lie superalgebras},
  year = {2001},
  journal = {Comment. Math. Helv.},
  volume = {76},
  pages = {110},
  doi = {10.1007/s000140050152},
  slaccitation = {
}

@article{Jimbo:1985ua,
  author = {Jimbo, Michio},
  title = {Quantum r Matrix for the Generalized Toda System},
  year = {1986},
  journal = {Commun. Math. Phys.},
  volume = {102},
  pages = {537-547},
  doi = {10.1007/BF01221646},
  slaccitation = {
}

@article{Jimbo:1985vd,
  author = {Jimbo, Michio},
  title = {A $q$ Analog of $u(Gl(n+1))$, Hecke Algebra and the Yang-Baxter Equation},
  year = {1986},
  journal = {Lett. Math. Phys.},
  volume = {11},
  pages = {247},
  doi = {10.1007/BF00400222},
  slaccitation = {
}

@article{Jimbo:1985zk,
  author = {Jimbo, Michio},
  title = {A q difference analog of U(g) and the Yang-Baxter equation},
  year = {1985},
  journal = {Lett. Math. Phys.},
  volume = {10},
  pages = {63-69},
  doi = {10.1007/BF00704588},
  slaccitation = {
}

@article{Jones,
  author = {Jones, V. F. R.},
  title = {Baxterization},
  year = {1991},
  journal = {Int. J. Mod. Phys.},
  volume = {A6},
  pages = {2035},
  doi = {10.1142/S0217751X91001027},
}

@article{kulish,
  author = {Kulish, P. P. and Reshetikhin, N. {\relax Yu}.},
  title = {Quantum linear problem for the Sine-Gordon andhigher representations},
  year = {1983},
  journal = {J. Sov. Math.},
  volume = {23},
  pages = {2435},
  warnincomplete = {WARNING},
}

@article{liebwu,
  author = {Lieb, E. H. and Wu, F. Y.},
  title = {Absence of Mott transition in an exact solution of the short-range, one-band model in one dimension},
  year = {1968},
  journal = {Phys. Rev. Lett.},
  volume = {20},
  pages = {1445-1448},
  doi = {10.1103/PhysRevLett.20.1445},
  slaccitation = {
}

@article{Maldacena:1998re,
  author = {Maldacena, Juan M.},
  title = {The large N limit of superconformal field theories and supergravity},
  year = {1998},
  journal = {Adv. Theor. Math. Phys.},
  volume = {2},
  pages = {231-252},
  archiveprefix = {arXiv},
  eprint = {hep-th/9711200},
  slaccitation = {
}

@article{Matsumoto:2007rh,
  author = {Matsumoto, Takuya and Moriyama, Sanefumi and Torrielli, Alessandro},
  title = {A Secret Symmetry of the AdS/CFT S-matrix},
  year = {2007},
  journal = {JHEP},
  volume = {0709},
  pages = {099},
  doi = {10.1088/1126-6708/2007/09/099},
  archiveprefix = {arXiv},
  eprint = {0708.1285},
  primaryclass = {hep-th},
  slaccitation = {
}

@article{Matsumoto:2009rf,
  author = {Matsumoto, Takuya and Moriyama, Sanefumi},
  title = {Serre Relation and Higher Grade Generators of the AdS/CFT Yangian Symmetry},
  year = {2009},
  journal = {JHEP},
  volume = {0909},
  pages = {097},
  doi = {10.1088/1126-6708/2009/09/097},
  archiveprefix = {arXiv},
  eprint = {0902.3299},
  primaryclass = {hep-th},
  slaccitation = {
}

@article{Nahm:1977tg,
  author = {Nahm, W.},
  title = {Supersymmetries and their representations},
  year = {1978},
  journal = {Nucl. Phys.},
  volume = {B135},
  pages = {149},
  doi = {10.1016/0550-3213(78)90218-3},
  slaccitation = {
}

@article{Perk_Schultz,
  author = {Perk, J. H. H. and Schultz, C. L.},
  title = {New families of commuting transfer matrices in $q$-state vertex models},
  year = {1981},
  journal = {Phys. Lett.},
  volume = {A84},
  pages = {407},
  doi = {10.1016/0375-9601(81)90994-4},
}

@article{Plefka:2006ze,
  author = {Plefka, Jan and Spill, Fabian and Torrielli, Alessandro},
  title = {On the Hopf algebra structure of the AdS/CFT S-matrix},
  year = {2006},
  journal = {Phys. Rev.},
  volume = {D74},
  pages = {066008},
  doi = {10.1103/PhysRevD.74.066008},
  archiveprefix = {arXiv},
  eprint = {hep-th/0608038},
  slaccitation = {
}

@article{qtm,
  author = {Juttner, G. and Kl\"umper, A. and Suzuki, J.},
  title = {The Hubbard chain at finite temperatures: ab initio calculations of Tomonaga-Luttinger liquid properties},
  year = {1998},
  journal = {Nucl. Phys.},
  volume = {B522},
  pages = {471},
  doi = {10.1016/S0550-3213(98)00256-9},
  archiveprefix = {arXiv},
  eprint = {cond-mat/9711310},
}

@article{Rej:2005qt,
  author = {Rej, Adam and Serban, Didina and Staudacher, Matthias},
  title = {Planar $\mathcal{N}$ = 4 gauge theory and the Hubbard model},
  year = {2006},
  journal = {JHEP},
  volume = {0603},
  pages = {018},
  doi = {10.1088/1126-6708/2006/03/018},
  archiveprefix = {arXiv},
  eprint = {hep-th/0512077},
  slaccitation = {
}

@article{Reshetikhin,
  author = {Reshetikhin, N.},
  title = {Multiparameter Quantum Groups and Twisted Quasitriangular Hopf Algebras},
  year = {1990},
  journal = {Lett. Math. Phys.},
  volume = {20},
  pages = {331},
  doi = {10.1007/BF00626530},
}

@article{Schultz,
  author = {Schultz, C. L.},
  title = {Solvable $q$-state models in lattice statistics and quantum field theory},
  year = {1981},
  journal = {Phys. Rev. Lett.},
  volume = {46},
  pages = {629},
  doi = {10.1103/PhysRevLett.46.629},
}

@article{shastry,
  author = {Shastry, B. S.},
  title = {Exact Integrability of the One-Dimensional Hubbard Model},
  year = {1986},
  journal = {Phys. Rev. Lett.},
  volume = {56},
  pages = {2453-2455},
  doi = {10.1103/PhysRevLett.56.2453},
  slaccitation = {
}

@article{Spill:2008tp,
  author = {Spill, Fabian and Torrielli, Alessandro},
  title = {On Drinfeld's second realization of the AdS/CFT su(2$/$2) Yangian},
  year = {2009},
  journal = {J. Geom. Phys.},
  volume = {59},
  pages = {489-502},
  doi = {10.1016/j.geomphys.2009.01.001},
  archiveprefix = {arXiv},
  eprint = {0803.3194},
  primaryclass = {hep-th},
  slaccitation = {
}

@article{Staudacher:2004tk,
  author = {Staudacher, Matthias},
  title = {The factorized S-matrix of CFT/AdS},
  year = {2005},
  journal = {JHEP},
  volume = {0505},
  pages = {054},
  doi = {10.1088/1126-6708/2005/05/054},
  archiveprefix = {arXiv},
  eprint = {hep-th/0412188},
  slaccitation = {
}

@article{Sutherland:1975vr,
  author = {Sutherland, Bill},
  title = {Model for a multicomponent quantum system},
  year = {1975},
  journal = {Phys. Rev.},
  volume = {B12},
  pages = {3795-3805},
  doi = {10.1103/PhysRevB.12.3795},
  slaccitation = {
}

@article{TJ1,
  author = {Brinkman, W. F. and Rice, T. M.},
  title = {Single-particle excitations in magnetic insulators},
  year = {1970},
  journal = {Phys. Rev.},
  volume = {B2},
  pages = {1324},
  doi = {10.1103/PhysRevB.2.1324},
}

@article{TJ2,
  author = {Lai, C. K.},
  title = {Lattice gas with nearest-neighbor interaction in one dimension with arbitrary statistics},
  year = {1974},
  journal = {J. Math. Phys.},
  volume = {15},
  pages = {1675},
  doi = {10.1063/1.1666522},
  slaccitation = {
}

@article{TJ4,
  author = {Essler, F. H. L. and Korepin, V. E.},
  title = {Higher conservation laws and algebraic Bethe ansatz for the supersymmetric $t$-$J$ model},
  year = {1992},
  journal = {Phys. Rev.},
  volume = {B46},
  pages = {9147},
  doi = {10.1103/PhysRevB.46.9147},
}

@article{Torrielli:2007mc,
  author = {Torrielli, Alessandro},
  title = {Classical r-matrix of the su(2$/$2) SYM spin-chain},
  year = {2007},
  journal = {Phys. Rev.},
  volume = {D75},
  pages = {105020},
  doi = {10.1103/PhysRevD.75.105020},
  archiveprefix = {arXiv},
  eprint = {hep-th/0701281},
  slaccitation = {
}

@article{Torrielli:2010kq,
  author = {Torrielli, Alessandro},
  title = {Review of AdS/CFT Integrability, Chapter VI.2: Yangian Algebra},
      journal        = "Lett.Math.Phys.",
      volume         = "99",
      pages          = "547-565",
      doi            = "10.1007/s11005-011-0491-z",
      year           = "2012",
  archiveprefix = {arXiv},
  eprint = {1012.4005},
  primaryclass = {hep-th},
  slaccitation = {
  warnincomplete = {WARNING},
}

@article{Yamane,
  author = {H. Yamane},
  title = {A central extension of $U_q(sl(2|2)^{(1)})$ and $R$-matrices with a new parameter},
  year = {2003},
  journal = {J. Math. Phys.},
  volume = {11},
  pages = {5450-5455},
  doi = {10.1063/1.1616251},
  archiveprefix = {arXiv},
  eprint = {math.QA/0304406},
  slaccitation = {
}

@article{Arutyunov:2009mi,
  author = {Arutyunov, Gleb and de Leeuw, Marius and Torrielli, Alessandro},
  title = {The Bound State S-Matrix for AdS$_5$ $\times$ S$^5$ Superstring},
  year = {2009},
  journal = {Nucl. Phys.},
  volume = {B819},
  pages = {319-350},
  doi = {10.1016/j.nuclphysb.2009.03.024},
  archiveprefix = {arXiv},
  eprint = {0902.0183},
  primaryclass = {hep-th},
  slaccitation = {
}

@article{Arutyunov:2009ce,
  author = {Arutyunov, Gleb and de Leeuw, Marius and Torrielli, Alessandro},
  title = {Universal blocks of the AdS/CFT Scattering Matrix},
  year = {2009},
  journal = {JHEP},
  volume = {0905},
  pages = {086},
  doi = {10.1088/1126-6708/2009/05/086},
  archiveprefix = {arXiv},
  eprint = {0903.1833},
  primaryclass = {hep-th},
  slaccitation = {
}

@article{Murgan:2008zu,
  author = {Murgan, Rajan and Nepomechie, Rafael I.},
  title = {q-deformed su(2$/$2) boundary S-matrices via the ZF algebra},
  year = {2008},
  journal = {JHEP},
  volume = {0806},
  pages = {096},
  doi = {10.1088/1126-6708/2008/06/096},
  archiveprefix = {arXiv},
  eprint = {0805.3142},
  primaryclass = {hep-th},
  slaccitation = {
}

@article{Hofman:2006xt,
  author = {Hofman, Diego M. and Maldacena, Juan Martin},
  title = {Giant magnons},
  year = {2006},
  journal = {J. Phys.},
  volume = {A39},
  pages = {13095-13118},
  doi = {10.1088/0305-4470/39/41/S17},
  archiveprefix = {arXiv},
  eprint = {hep-th/0604135},
  slaccitation = {
}

@article{Galleas:2009ye,
  author = {Galleas, W.},
  title = {The Bethe Ansatz Equations for Reflecting Magnons},
  year = {2009},
  journal = {Nucl. Phys.},
  volume = {B820},
  pages = {664-681},
  doi = {10.1016/j.nuclphysb.2009.04.024},
  archiveprefix = {arXiv},
  eprint = {0902.1681},
  primaryclass = {hep-th},
  slaccitation = {
}

@article{Correa:2008av,
  author = {Correa, D. H. and Young, C. A. S.},
  title = {Reflecting magnons from D7 and D5 branes},
  year = {2008},
  journal = {J. Phys.},
  volume = {A41},
  pages = {455401},
  doi = {10.1088/1751-8113/41/45/455401},
  archiveprefix = {arXiv},
  eprint = {0808.0452},
  primaryclass = {hep-th},
  slaccitation = {
}

@article{deLeeuw:2010nd,
  author = {de Leeuw, Marius},
  title = {The S-matrix of the AdS$_5$ $\times$ S$^5$ superstring},
  archiveprefix = {arXiv},
  eprint = {1007.4931},
  primaryclass = {hep-th},
  slaccitation = {
  note = {PhD Thesis},
  warnincomplete = {WARNING},
}

@article{Zhang:2004qx,
  author = {Zhang, Yao-Zhong and Gould, Mark D.},
  title = {A Unified and Complete Construction of All Finite Dimensional Irreducible Representations of gl(2$/$2)},
  year = {2005},
  journal = {J. Math. Phys.},
  volume = {46},
  pages = {013505},
  doi = {10.1063/1.1812829},
  archiveprefix = {arXiv},
  eprint = {math/0405043},
  slaccitation = {
}

@article{Delius:1994ab,
  author = {Delius, Gustav W. and Gould, Mark D. and Links, Jon R. and Zhang, Yao-Zhong},
  title = {Solutions of the Yang-Baxter equation with extra nonadditive parameters. 2: U$_q$(gl(m$/$n))},
  year = {1995},
  journal = {J. Phys.},
  volume = {A28},
  pages = {6203-6210},
  doi = {10.1088/0305-4470/28/21/023},
  archiveprefix = {arXiv},
  eprint = {hep-th/9411241},
  slaccitation = {
}

@Article{Arutyunov:2009ga,
     author    = "Arutyunov, Gleb and Frolov, Sergey",
     title     = "{Foundations of the AdS$_5$ $\times$ S$^5$ Superstring. Part I}",
     journal   = "J. Phys.",
     volume    = "A42",
     year      = "2009",
     pages     = "254003",
     eprint    = "0901.4937",
     archivePrefix = "arXiv",
     primaryClass  =  "hep-th",
     doi       = "10.1088/1751-8113/42/25/254003",
     SLACcitation  = "
}

@Article{Chen:2006gp,
     author    = "Chen, Heng-Yu and Dorey, Nick and Okamura, Keisuke",
     title     = "{The asymptotic spectrum of the $\mathcal{N}$ = 4 super Yang-Mills spin
                  chain}",
     journal   = "JHEP",
     volume    = "03",
     year      = "2007",
     pages     = "005",
     eprint    = "hep-th/0610295",
     archivePrefix = "arXiv",
     doi       = "10.1088/1126-6708/2007/03/005",
     SLACcitation  = "
}

\end{bibtex}

\pdfbookmark[1]{\refname}{references}
\bibliography{\jobname}
\bibliographystyle{nb}

\begin{mpostcmd}
verbatimtex 
\end{document}